\documentclass[aps,prd,amssymb,nofootinbib,twocolumn,epsf,floatfix,showpacs]
{revtex4}
\usepackage[usenames]{color} 
\usepackage{ulem} 
\usepackage{graphicx}
\usepackage{amssymb}
\usepackage{amsmath}
\usepackage{epstopdf}

\newcommand{\half}{{\tfrac{1}{2}}}
\newcommand{\fourth}{{\tfrac{1}{4}}}

\begin{document}
 

\title{Solving Einstein's equation numerically on manifolds\\ with arbitrary
       spatial topologies}

\author{Lee Lindblom, B\'ela Szil\'agyi, and Nicholas W. Taylor}

\affiliation{Theoretical Astrophysics 350-17, California Institute of
Technology, Pasadena, California 91125, USA}

\date{\today}
 
\begin{abstract} 
This paper develops a method for solving Einstein's equation
numerically on multicube representations of manifolds with arbitrary
spatial topologies.  This method is 
designed to provide a set of
flexible, easy to use computational procedures 
that make it possible to explore 
the never before studied properties of    
solutions to Einstein's equation on manifolds with arbitrary 
toplogical structures.
A new covariant, first-order symmetric-hyperbolic
representation of Einstein's equation is developed for this purpose,
along with the needed boundary conditions 
at the
interfaces between adjoining cubic regions. 
Numerical tests are
presented that demonstrate the long-term numerical stability of this
method for evolutions of a complicated, time-dependent solution 
of Einstein's
equation coupled to a complex scalar field 
on a manifold with
spatial topology $S^3$.  The accuracy of these numerical test
solutions is 
evaluated by performing convergence studies and by
comparing the full nonlinear numerical 
results to the analytical
perturbation 
solutions, which are also derived here.
\end{abstract}
 
\maketitle

\section{Introduction}
\label{s:Introduction}

Solving partial differential equations on manifolds with arbitrary
spatial topologies presents 
a number of challenges beyond those
required to solve those equations on subsets of ${R}^3$.  In a
previous paper, Lindblom and Szil\'agyi~\cite{LindblomSzilagyi2011a}
showed how systems of elliptic and hyperbolic partial differential
equations for collections of tensor fields can be solved numerically
on manifolds with arbitrary spatial topologies by using multicube
representations of those manifolds.  We review some of the basic
features of that multicube method in
Sec.~\ref{s:ReviewMultiCubeMethod}.  In particular, we discuss how the
global differentiable structure (needed to define what it means
globally to have smooth tensor fields) can be defined conveniently for
multicube manifolds.  We also review what boundary conditions are
needed at the interfaces between cubic regions and how these
conditions are enforced for first-order symmetric-hyperbolic evolution
systems.

In Sec.~\ref{s:CovariantEinsteinSystem} we develop a new
(spatially) covariant, first-order symmetric-hyperbolic representation
of the Einstein system that can be used on manifolds with arbitrary
spatial topologies.  The standard generalized-harmonic representation
of Einstein's equation~\cite{Lindblom2006} is a special
case of these new covariant representations on manifolds whose spatial
slices are subsets of $R^3$.  Given this new representation of the
Einstein system, it is straightforward to adapt the multicube methods
developed by 
Lindblom and Szil\'agyi~\cite{LindblomSzilagyi2011a} to
the Einstein case.  In particular, the explicit boundary conditions
that must be applied to the characteristic fields of this system at
the interface boundaries between adjoining cubic regions are presented
in Sec.~\ref{s:CovariantEinsteinSystem}.

The long-term numerical stability of these methods is 
tested in
Secs.~\ref{s:EinsteinStatic}--\ref{s:PerturbedEinsteinStatic} by
studying solutions to Einstein's equation coupled to a complex
Klein-Gordon scalar field.  There exists a static solution to this
system of equations whose spatial geometry is the standard round
metric on $S^3$.  This solution is therefore 
a (new) representation of
the Einstein static universe.  The Einstein static universe has a 
well-known physical instability that causes the universe to expand without
bound or to collapse to a singularity on a fairly short time scale
(cf. Ref.~\cite{Eddington1030}). 
Our numerical tests of the coupled Einstein-Klein-Gordon system,
described in Sec.~\ref{s:EinsteinStatic}, reproduce this well-known
result.

One important goal of this paper is to study the long-term numerical
stability of our implementation of the multicube methods.  Since the
Einstein-Klein-Gordon static universe solution is unstable, we
introduce unphysical mode-damping forces 
into the Einstein and
Klein-Gordon equations that are designed to 
exponentially suppress the two unstable
modes of this solution.  
One of these unstable modes is the well-known
spatially homogeneous physical instability of the Einstein static
universe, while the other is a dipole instability that exists in the
particular coordinate gauge used in our tests.  These 
mode-damping forces, described in detail in Sec.~\ref{s:ModeDamping}, leave
untouched all of the rich dynamics of the Einstein-Klein-Gordon
evolution equations, except for the  
degrees of freedom associated
with the unstable modes.  
With the addition of these mode-damping forces, we are able to
perform long-term evolutions (about 160 light-crossing times) of the
Einstein static universe.  The results of these tests, described in
Sec.~\ref{s:ModeDamping}, show that our implementation of the
multicube method is stable and convergent, even on 
such very long time scales.  
We show that the constraints of this system, as well as
the unphysical mode-damping forces, converge (exponentially quickly)
toward zero as the spatial resolution of the 
numerical solutions is increased.

Finally, we test the accuracy and numerical stability of our
implementation of the multicube method in
Sec.~\ref{s:PerturbedEinsteinStatic} by studying a complicated,
time-dependent solution of the coupled Einstein-Klein-Gordon system.
We derive the general solution to these equations analytically for
first-order perturbations of the Einstein-Klein-Gordon static universe
solution.  These analytical solutions are then used to construct
initial data composed of a superposition of 15 distinct modes.
We evolve these initial data numerically and 
demonstrate stability and convergence.  We show 
that the constraints of the system 
and the magnitudes of the unphysical mode-damping forces 
converge exponentially toward zero as the
spatial resolution
is increased.  We measure the accuracy of the
numerical solutions by comparing them with the analytical first-order
perturbation solutions.  
We show that the differences
between these two solutions converge toward zero, until 
these differences reach
the level of the neglected quadratic terms in the analytical
perturbation solution.  These accuracy and stability tests are carried
out for this complicated time-dependent solution for about 160 
light-crossing times of the solution.

Solving Einstein's equation 
numerically on manifolds with arbitrary
spatial topologies requires a number of computational tools beyond
those needed to solve problems on manifolds having spatial slices
which can be embedded in $R^3$.  In particular, smooth tensor fields
must be represented in a way that does not depend on the existence of
a single, smooth global coordinate system.  To our knowledge, the
methods developed by Lindblom and
Szil\'agyi~\cite{LindblomSzilagyi2011a} and applied here to Einstein's
equation are the first numerical methods to appear in the
literature that are capable of solving these equations on aribitrary
manifolds.  
As far as we know,
Bentivegna and Korzynski~\cite{Bentivegna2012,
  Bentivegna2013, Bentivegna2013a} give the only other published
results of fully three-dimensional numerical solutions of Einstein's
equations on manifolds with nontrivial topologies. 
They evolve Einstein's equation in vacuum on manifolds having
spatial topologies $S^3$ and $T^3$, with 
black hole lattice solutions. 
They avoid the generic problem of solving equations on
manifolds with aribitrary topologies by embedding each of their
spatial manifolds\footnote{One of the black hole interiors in
the Bentivegna and Korzynski $S^3$ solution is excised, and a 
conformal transformation
is applied to map its horizon to infinity in $R^3$.}
in $R^3$ and using its global Cartesian coordinates
to represent smooth tensors.  They then solve Einstein's equation
numerically in $R^3$ 
using the standard tools of
numerical relativity.

\section{Review of the MultiCube Method}
\label{s:ReviewMultiCubeMethod}

The most useful manifolds for solving Einstein's equation numerically
are those which admit globally hyperbolic causal structures.  These
manifolds have topologies of the form $R\times\Sigma$, where $\Sigma$
is a three-dimensional manifold.  The multicube method of
representing three-dimensional manifolds with arbitrary topologies
consists of three basic elements: (i) a collection of
nonoverlapping cubic blocks ${\cal B}_A$ that cover the manifold,
(ii) a collection of maps $\Psi^{A\alpha}_{B\beta}$ that specify
how the faces of the blocks are connected together to create the desired
topology, and (iii) a smooth positive-definite reference metric
$\tilde g_{ij}$ used to determine the differentiable structure of the
manifold.  We devote most of the remainder of this section to a
discussion of these basic elements of the multicube method.  
In addition, we give a brief review of the interface boundary conditions
needed to solve first-order symmetric-hyperbolic evolution systems,
like Einstein's equation, on multicube manifolds.

\subsection{Multicube structures}
\label{s:MultiCubeStructures}

An arbitrary (three-dimensional) manifold $\Sigma$ can be subdivided
into a collection of regions, each of which can be mapped smoothly
into a cube ${\cal B}_A$ in $R^3$ (cf. Ref.~\cite{LindblomSzilagyi2011a}).  
We use upper-case latin
indices $\scriptstyle \{A,B,...\}$ with
$\scriptstyle{A}=\{1,2,...,N\}$ to label these regions and their
images ${\cal B}_A$ in $R^3$.  These regions overlap in $\Sigma$ only
along the boundaries between neighboring regions.  
It is convenient to
choose the images of these regions ${\cal B}_A$ 
to be cubes
of uniform coordinate size, $L$, which are all oriented along the same
global Cartesian coordinate axes in $R^3$.  In this case the cube
${\cal B}_A$ can be specified simply by giving the location of its
center $\vec c_A=(c{}^x{}_A,c{}^y{}_A,c{}^z{}_A)$ in $R^3$.  It is
also convenient to arrange the cubes ${\cal B}_A$ so they intersect
(if at all) in $R^3$ only at points on faces where the corresponding
regions touch in $\Sigma$.

This collection of cubes ${\cal B}_A$ provides the basic framework on
which a multicube representation of the manifold $\Sigma$ can be
constructed.  Each point in the interior of one of the 
cubes represents a unique point in $\Sigma$.  
In addition, each
point in $\Sigma$ is the inverse image of at least one point in 
the closure of
$\cup_A{\cal B}_A$. The Cartesian coordinates of $R^3$ therefore
provide a global way of identifying points in $\Sigma$.  We use the
notation $x^i=\{x,y,z\}$ to denote these coordinates, where latin
indices $\scriptstyle \{i,j,k,\ell,...\}$ are used to denote spatial
quantities.

\subsection{Interface boundary maps}
\label{s:InterfaceBoundaryMaps}

The topological structure of the manifold $\Sigma$ determines how the
cubic regions ${\cal B}_A$ are connected together.  Conversely, the
topological structure of a multicube manifold is determined by giving
a collection of maps $\Psi^{A\alpha}_{B\beta}$ that specify how the
points on the faces of each cubic region are identified with those of
its neighbors~\cite{LindblomSzilagyi2011a}.  We use the notation
$\Psi^{A\alpha}_{B\beta}$ to represent the map from 
the $\partial_\alpha{\cal B}_A$ face of cube ${\cal B}_A$ to 
the $\partial_\beta{\cal B}_B$ face of cube ${\cal
  B}_B$.  We use lower-case greek indices $\scriptstyle\{
\alpha,\beta,...\}$ with $\scriptstyle \alpha=\{\pm x,\pm y, \pm z\}$
to label the faces of each cube.
The cubes $\mathcal{B}_A$ are chosen to be aligned with the
global Cartesian coordinate axes in $R^3$, so the 
region boundary faces are 
always located at constant spatial coordinate surfaces.
For example,
the boundary $\partial_\alpha {\cal B}_A$ is assumed to be a surface
of constant coordinate 
$x_A^{\sigma}=x_A^{|\alpha|}$, where the
index $\sigma=|\alpha|$ 
denotes the fixed 
boundary-surface
coordinate.  This boundary surface is identified with the boundary
$\partial_\beta {\cal B}_B$, a surface of constant coordinate
$x_B^{\sigma}=x_B^{|\beta|}$, via the map $\Psi^{A\alpha}_{B\beta}$.
 
The map $\Psi^{A\alpha}_{B\beta}$ that takes the Cartesian coordinates
$x_B^j$ of points in $\partial_\beta{\cal B}_B$ to the Cartesian
coordinates $x_A^i$ of points in $\partial_\alpha{\cal B}_A$ can be
chosen to have the form of a simple translation plus 
rotation and/or reflection (cf. Ref.~\cite{LindblomSzilagyi2011a}):
\begin{eqnarray}
x_A^i&=&c_A^i+f_\alpha^i+C^{A\alpha i}_{B\beta j}\,\big(x_B^j 
- c_B^j - f_\beta^j\big).
\end{eqnarray}
The vector $c_A^i+f_\alpha^i$ is the location of the center of the
face $\partial_\alpha{\cal B}_A$, and $C^{A\alpha i}_{B\beta j}$ is
the combined spatial rotation and reflection matrix needed to match
the face $\partial_\alpha{\cal B}_A$ to the face $\partial_\beta{\cal
  B}_B$ in the desired way.  The vectors $c_A^i+f_\alpha^i$ and
matrices $C^{A\alpha i}_{B\beta j}$ in these maps are 
constants determined once and for all by the topology of the particular
manifold.  These maps are smooth for the coordinates $x^k$ within the
boundary surface, i.e., for those with 
$k\neq\sigma$. For the normal surface coordinate
$x^\sigma$, however, the maps are only continuous and not (in general)
differentiable.

The multicube Cartesian coordinates $x_A^i$ on the 3-manifold
$\Sigma$ can be extended naturally to coordinates on the spacetime
$R\times\Sigma$: $x_A^a=\{t_A,x_A^i\}$, where latin indices from the
beginning of the alphabet, $\scriptstyle\{ a,b,...\}$ with
$\scriptstyle a=\{t,x,y,z\}$, denote spacetime quantities.  The maps
$\Psi^{A\alpha}_{B\beta}$ defined above can be extended in a natural
way to include the equation for the continuity of the time coordinate
across region boundaries, $t_A=t_B$.  The full spacetime coordinate
transformation map can then be written in the compact, 
four-dimensional notation
\begin{eqnarray}
x_A^a&=&c_A^a+f_\alpha^a+C^{A\alpha a}_{B\beta b}\,\big(x_B^b 
- c_B^b - f_\beta^b\big),
\label{e:SurfaceCoordinateMap}
\end{eqnarray}
where $c_A^t+f_\alpha^t=0$, $C^{A\alpha t}_{B\beta b}=\delta^t_b$, and
$C^{A\alpha a}_{B\beta t}=\delta^a_t$.

Explicit expressions for the multicube representations of the
  3-manifolds $T^3$, $S^1\times S^2$, and $S^3$ are described in
  detail in Ref.~\cite{LindblomSzilagyi2011a}.  In particular,
  specific expressions are given there for the collections of cubic
  regions $\mathcal{B}_A$, the vectors $c_A^i$ and $f_\alpha^i$, and the
  interface boundary transformation matrices $C^{A\alpha i}_{B\beta
    j}$, needed to construct the multicube representation of each of
  these manifolds.

\subsection{Reference metrics}
\label{s:ReferenceMetric}

Tensor fields can be represented on multicube manifolds by giving
their components (expressed in the global coordinate basis of $R^3$)
as functions of the global Cartesian coordinates.  Within each
coordinate region ${\cal B}_A$, the components of smooth tensor
fields are smooth functions of these coordinates $x_A^a$.  Additional
structure must be provided, however, 
that determines how to transform continuous,
differentiable, and smooth tensor fields 
across the interface boundaries between regions in multicube manifolds.  
One way
to fix this differentiable structure is to specify a smooth, static
spacetime metric, which we denote as $\tilde \psi_{ab}$
(cf. Ref.~\cite{LindblomSzilagyi2011a}).  Like other smooth vector and
tensor fields, the components of $\tilde \psi_{ab}$ 
might be discontinuous
across the boundaries of the cubic block regions when written in terms
of the global multicube Cartesian coordinate basis.  However, the
components of $\tilde \psi_{ab}$ 
must be smooth functions in any smooth atlas of overlapping
coordinate charts. The numerical examples studied in this paper solve
Einstein's equation on a manifold with the topology of a
three-sphere, $\Sigma=S^3$.  For these examples, the multicube
representation of the standard round-sphere metric on $S^3$ can be
used to construct a reference metric
(cf. Ref.~\cite{LindblomSzilagyi2011a}).  Smooth multicube reference
metrics are also given in Ref.~\cite{LindblomSzilagyi2011a}
for manifolds with spatial topologies $T^3$ and $S^1\times S^2$.
In a future paper we will
describe an algorithm for constructing smooth reference metrics
$\tilde \psi_{ab}$ on any multicube manifold.

It is easy to construct covectors that are normal to the boundaries
of the multicube regions: $\tilde n_{Aa}\propto\partial_a
x^\sigma_A$.  Given a smooth reference metric $\tilde \psi_{ab}$,
these covectors can be normalized to be outward pointing and to have
unit length: $\tilde n_A^a\tilde n_A^b\tilde \psi_{ab}=1$ and $\tilde
n_{Aa}=\tilde \psi_{ab}\tilde n^b_A$.  Let $\tilde n_A^a$ denote the
outward-directed unit normal to the boundary $\partial_\alpha{\cal
  B}_A$, and $\tilde n_B^a$ the outward-directed unit normal to
$\partial_\beta{\cal B}_B$.  Since the reference metric $\tilde
\psi_{ab}$ is smooth, these normal vectors (up to sign) represent the 
same vector at the corresponding points on each side of identified
boundaries.  The transformation law that maps smooth tensor fields
across interface boundaries must therefore be constructed to transform
$\tilde n_B^a$ into $-\tilde n_A^a$.  In contrast, continuous vector
fields $u^a_A$ that are tangent to the boundary, i.e., $u^a_A\tilde
n^b_A\tilde \psi_{ab}=0$, should transform using the standard Jacobian
of the map $\Psi^{A\alpha}_{B\beta}$ in
Eq.~(\ref{e:SurfaceCoordinateMap}): $u^a_A=C^{A\alpha a}_{B\beta
  b}u^b_B$.  It is straightforward then to construct the
transformations, effectively Jacobians, needed to transform arbitrary
tensor fields from the region boundary $\partial_\beta{\cal B}_B$ to
$\partial_\alpha{\cal B}_A$:
\begin{eqnarray}
J^{A\alpha a}_{B\beta b} &=& C^{A\alpha a}_{B\beta c}
(\delta^c_b - \tilde n_B^c \tilde n_{Bb} ) - \tilde n_A^a \tilde n_{Bb},\\
J_{A\alpha a}^{*B\beta b} &=& 
(\delta_a^c - \tilde n_{Aa} \tilde n_A^c )C_{A\alpha c}^{B\beta b}
 - \tilde n_{Aa} \tilde n_B^b. 
\end{eqnarray}
These effective Jacobians transform the background surface normals correctly,
\begin{eqnarray}
\tilde n^a_A&=&-J^{A\alpha a}_{B\beta b}\tilde n_B^b,\\
\tilde n_{Aa} &=&-J_{A\alpha a}^{*B\beta b}\tilde n_{Bb},
\end{eqnarray}
and they also transform the components of vectors 
$u^a$ that are tangent to the boundary correctly,
\begin{eqnarray}
u^a_A&=&J^{A\alpha a}_{B\beta b}u_B^b=\,\,C^{A\alpha a}_{B\beta b}u_B^b,
\end{eqnarray}
using the rotation/reflection matrix $C^{A\alpha a}_{B\beta b}$ from
the surface coordinate map.  The Jacobian and its dual are also
inverses of one another:
\begin{eqnarray}
\delta^{Aa}_{Ab} = J^{A\alpha a}_{B\beta c}J^{*B\beta c}_{A\alpha b}.
\end{eqnarray}

We introduce the notation $\langle v_B^a\rangle_A$ and $\langle
w_{Ba}\rangle_A$ to denote the result of transforming these vector and
covector fields from the boundary of region $\scriptstyle B$ to the
corresponding points on the boundary of region $\scriptstyle A$:
\begin{eqnarray}
\langle v^a_B\rangle_A&=&J^{A\alpha a}_{B\beta b}v_B^b
\label{e:FieldTransformationA},\\
\langle w_{Ba}\rangle_A &=&J_{A\alpha a}^{*B\beta b}w_{Bb}.
\label{e:FieldTransformationB}
\end{eqnarray}
The necessary and sufficient conditions for the continuity of these
fields across interface boundaries are $v_A^a = \langle
v_B^a\rangle_A$ and $w_{Aa}=\langle w_{Ba}\rangle_A$.  The appropriate
transformation laws for tensor fields are obtained by applying the
effective Jacobian to each index of the tensor.  For example, the
physical spacetime metric $\psi_{ab}$, which will 
generally be different than the static reference metric $\tilde\psi_{ab}$, 
transforms across interface
boundaries as follows:
\begin{eqnarray}
\langle\psi_{Bab}\rangle_A = 
J^{*B\beta c}_{A\alpha a} J^{*B\beta d}_{A\alpha b}\psi_{Bcd}.
\end{eqnarray}
The continuity of the spacetime metric across this boundary is 
the statement that $\psi_{Aab} = \langle\psi_{Bab}\rangle_A$.

The rules for transforming the derivatives of tensors across interface
boundaries can be determined by introducing the covariant derivative
$\tilde \nabla_a$ that is compatible with the smooth reference metric,
i.e., $\tilde \nabla_c \tilde \psi_{ab}=0$.  The covariant
derivatives of smooth tensors are tensors, so these derivatives are
transformed across region boundaries using the effective Jacobian
$J^{A\alpha a}_{B\beta b}$ defined above.  In particular, the
transformations of the covariant derivatives of the vector $v^a$ and
covector $w_a$ are given by the expressions
\begin{eqnarray}
\langle\tilde \nabla_{a} v_B^b\rangle_A 
&=& J^{*B\beta c}_{A\alpha a} J_{B\beta d}^{A\alpha b}
\tilde \nabla_{c} v_B^d,\nonumber\\
\langle\tilde \nabla_{a} w_{Bb}\rangle_A 
&=& J^{*B\beta c}_{A\alpha a} J^{*B\beta d}_{A\alpha b}
\tilde \nabla_{c} w_{Bd}.\nonumber
\end{eqnarray}
Tensor fields with continuous derivatives therefore satisfy the
continuity conditions $\tilde \nabla_{a}v_A^b =\langle \tilde
\nabla_{a} v_B^b\rangle_A $ and $\tilde \nabla_{a}w_{Ab}=\langle\tilde
\nabla_{a}w_{Bb}\rangle_A$. These transformation laws can be
generalized to tensor fields of arbitrary rank in the obvious way. 
In particular, the transformation of the derivatives of the
spacetime metric is given by
\begin{eqnarray}
\langle\tilde \nabla_{c}\psi_{Bab}\rangle_A &=& J^{*B\beta d}_{A\alpha c}
J^{*B\beta e}_{A\alpha a} J^{*B\beta f}_{A\alpha b}
\tilde \nabla_{d}\psi_{Bef}.\nonumber
\end{eqnarray}
Smooth tensor fields are defined to be those having continuous derivatives
of all orders.

\subsection{Boundary conditions for hyperbolic systems}
\label{s:BCHyperbolicSystems}

A first-order symmetric-hyperbolic system of equations for the
dynamical fields $u^{\cal A}$ (assumed here to be a collection of
tensor fields) can be written in the form
\begin{eqnarray}
\partial_t u^{\cal A} + A^{k{\cal A}}{}_{\cal B}(\mathbf{x},\mathbf{u})
\,\tilde\nabla_k u^{\cal
  B} = F^{\cal A}(\mathbf{x},\mathbf{u}),
\label{e:FirstOrderHyperbolicEq}
\end{eqnarray}
where the characteristic matrix, $A^{k{\cal A}}{}_{\cal
  B}(\mathbf{x},\mathbf{u})$, and the source term, $F^{\cal
  A}(\mathbf{x},\mathbf{u})$, may depend on the spacetime coordinates
$x^a$ and the fields $u^{\cal A}$, but not their derivatives.  The
script indexes $\scriptstyle \{ {\cal A}, {\cal B}, {\cal C}, ...\}$
in these expressions label the components 
of the collection of tensor
fields that make up $u^{\cal A}$. These systems are called symmetric
because, by assumption, there exists a positive-definite metric on the
space of fields, $S_{\cal AB}$, that can be used to transform the
characteristic matrix into a symmetric form: $S_{\cal AC}A^{k\,{\cal
    C}}{}_{\cal B}\equiv A^k_{\cal AB}=A^k_{\cal BA}$.

Boundary conditions for symmetric-hyperbolic systems must be imposed
on the incoming characteristic fields of the system.  The
characteristic fields $\hat u^{{\cal K}}$ (whose index 
${\scriptstyle {\cal K}}$ labels the collection of characteristic fields) are
projections of the dynamical fields $u^{\cal A}$ onto the matrix of left
eigenvectors of the characteristic matrix (cf. Refs.~\cite{Kidder2005,
  Lindblom2006}):
\begin{eqnarray}
\hat u^{{\cal K}} = e^{{\cal K}}{}_{\!{\cal A}}(\mathbf{n})\, u^{\cal A}.
\end{eqnarray}
The matrix of eigenvectors $e^{{\cal K}}{}_{\!{\cal A}}(\mathbf{n})$
is defined by the equation
\begin{eqnarray}
e^{{\cal K}}{}_{\!{\cal A}}(\mathbf{n})\,n_kA^{k\,{\cal A}}{}_{\cal B}(u)
= v_{({\cal K})}\,e^{{\cal K}}{}_{\!{\cal B}}(\mathbf{n}),
\end{eqnarray}
where the covector $n_k$ that appears in this definition is the
outward-pointing unit normal to the surface on which the
characteristic fields are evaluated. The eigenvalues $v_{({\cal K})}$
are often referred to as the characteristic speeds of the system.  The
characteristic fields $\hat u^{{\cal K}}$ represent the independent
dynamical degrees of freedom at the boundaries.  These characteristic
fields propagate at the speeds $v_{({\cal K})}$ (in the short
wavelength limit), so boundary conditions must be given for each
incoming characteristic field, i.e., for each field with speed
$v_{({\cal K})}<0$.  No boundary condition is required (or allowed)
for outgoing characteristic fields, i.e., for any field with 
$v_{({\cal K})}\geq0$.

The boundary conditions on the dynamical fields $u^{\cal A}$ that
ensure the equations are satisfied across the faces of adjoining cubic
regions are quite simple: data for the incoming characteristic fields
at the boundary of one region are supplied by the outgoing
characteristic fields from the neighboring region.  The boundary
conditions at an interface between cubic regions require that the
dynamical fields $u^{\cal A}_A$ in region ${\cal B}_A$ be transformed
into the representation used in the neighboring region ${\cal B}_B$.
When the dynamical fields $u^{\cal A}$ are a collection of tensor
fields (as assumed here), their components are transformed from one
coordinate representation to another using the Jacobians of the
transformation as described in Eqs.~(\ref{e:FieldTransformationA}) and
(\ref{e:FieldTransformationB}).  In this case, the needed boundary
conditions can be stated precisely for hyperbolic evolution problems:
Consider two cubic regions ${\cal B}_A$ and ${\cal B}_B$ whose
boundaries $\partial_\alpha {\cal B}_A$ and $\partial_\beta {\cal
  B}_B$ are identified by the map $\Psi^{\alpha A}_{\beta B}$ as
defined in Eq.~(\ref{e:SurfaceCoordinateMap}).  The required boundary
conditions on the dynamical fields $u^{\cal A}_A$ consist of fixing
the incoming characteristic fields $\hat u^{{\cal K}}_A$ (i.e., those
with speeds $v_{({{\cal K}})}<0$) at the boundary $\partial_\alpha
{\cal B}_A$ with data, $u^{\cal B}_B$, from the fields on the
neighboring boundary $\partial_\beta {\cal B}_B$:
\begin{eqnarray}
\hat u^{{\cal K}}_{{A}} &=& \langle e^{{\cal K}}{}_{\!\cal A}(\mathbf{n})\rangle_A
\langle u^{\cal A}_B\rangle_A.
\label{e:HyperbolicBC}
\end{eqnarray}
The matrix of eigenvectors, $\langle e^{{\cal K}}{}_{\!\cal
  A}(\mathbf{n})\rangle_A$, that appears in Eq.~(\ref{e:HyperbolicBC})
is to be constructed with the fields from region ${\cal B}_B$ that
have been transformed into region ${\cal B}_A$ where the boundary
condition is to be imposed.  This boundary condition must be applied
to each incoming characteristic field on each internal cube 
face---i.e., on each face that is identified with the face of a neighboring
region.

\section{Covariant First-Order Einstein Evolution System}
\label{s:CovariantEinsteinSystem}

Einstein's equation determines the spacetime metric $\psi_{ab}$ by
equating the Einstein curvature tensor to the stress-energy tensor of
the matter in the spacetime.  This equation is, of course, covariant.
The standard first-order hyperbolic representations of Einstein's
equation (e.g., Ref.~\cite{Lindblom2006}), however, are not covariant,
because the auxiliary dynamical fields introduced to make the system
first order are not tensors.  This lack of covariance has not caused
any problems (that we know of) in the codes that solve these
noncovariant equations on spatial manifolds that can be embedded in
$R^3$, e.g., for binary black-hole spacetimes.  However, our attempts
to use these noncovariant representations for numerical evolutions on
manifolds with nontrivial
spatial topologies failed.  We were 
unable to achieve stable and convergent evolutions, at the interface
boundaries in particular.  These problems 
disappeared when we adopted
the spatially covariant representation of the first-order Einstein
evolution system described in the remainder of this section.  The
interface boundary conditions needed for this new covariant
representation are precisely those described in
Sec.~\ref{s:BCHyperbolicSystems} for any hyperbolic system whose
dynamical fields are tensors.

Let $\psi_{ab}$ denote the physical spacetime metric that is
determined by solving Einstein's equation, and let $\Gamma^a_{bc}$ and
$\nabla_a$ denote the connection and covariant derivative associated
with $\psi_{ab}$.  Let $\tilde\psi_{ab}$ denote a smooth static
reference metric, and let $\tilde \Gamma^a_{bc}$ and $\tilde \nabla_a$
denote the connection and covariant derivative associated with
$\tilde\psi_{ab}$.  It is straightforward to show that the physical
Ricci curvature $R_{ab}$ associated with $\psi_{ab}$ satisfies the
identity
\begin{eqnarray}
R_{ab}&=& -{\half}\psi^{cd}\tilde\nabla_c\tilde\nabla_d\psi_{ab}
+\nabla_{(a}\Delta_{b)}-\psi^{cd}\tilde R^{e}{}_{cd(a}\psi_{b)e}
\nonumber\\
&& + \psi^{cd}\psi^{ef}\left(\tilde\nabla_e\psi_{ca}
\tilde\nabla_f\psi_{bd}-\Delta_{ace}\Delta_{bdf}\right),
\label{e:RicciIdentity}
\end{eqnarray}
where $\Delta_{abc}$ is the tensor that describes the difference
between the connections:
\begin{eqnarray}
\Delta_{abc}&=&
\psi_{ad}\left(\Gamma^d_{bc}-\tilde\Gamma^d_{bc}\right)\nonumber\\
&=&{\half}\left(
\tilde\nabla_b\psi_{ac}+\tilde\nabla_c\psi_{ab}-\tilde\nabla_a\psi_{bc}\right).
\label{e:DeltaDef}
\end{eqnarray}
The vector $\Delta_a$ is defined as $\Delta_a=\psi^{bc}\Delta_{abc}$,
and $\tilde R^{d}{}_{abc}$ is the reference Riemann curvature
associated with $\tilde\psi_{ab}$.  Note that
Eq.~(\ref{e:RicciIdentity}) reduces to Eq.~(4) of
Ref.~\cite{Lindblom2006} for the case where the reference metric is
the flat Minkowski metric $\tilde\psi_{ab}=\eta_{ab}$ expressed in
Cartesian coordinates.

In analogy with the generalized harmonic representations of Einstein's
equation (e.g., Ref.~\cite{Lindblom2006}), the gauge (or coordinate)
conditions are fixed in this covariant evolution system by setting
$\Delta_a$ to be a fixed gauge source function:
\begin{eqnarray}
\Delta_a=-H_a.
\end{eqnarray}  
We assume that this gauge source function $H_a=H_a(\psi,\tilde\psi,
\partial^k\tilde\psi,x)$ may depend on the physical metric $\psi_{ab}$
(but not its derivatives) and the reference metric $\tilde\psi_{ab}$
(including its derivatives if desired), as well as the spacetime
coordinates $x^a$.  This gauge condition becomes a constraint of
the system:
\begin{eqnarray}
{\cal C}_a = \Delta_a + H_a.
\label{e:ConstraintGauge}
\end{eqnarray}
The covariant vacuum evolution equation therefore satisfies the
standard generalized harmonic evolution equation:
\begin{eqnarray}
0=R_{ab} - \nabla_{(a}{\cal C}_{b)}. 
\end{eqnarray}
The standard argument (cf. Ref.~\cite{Lindblom2006})
using the Bianchi identities 
implies that the constraint
${\cal C}_a$ satisfies the evolution equation
\begin{eqnarray}
0=\nabla^b\nabla_b{\cal C}_a +{\cal C}^b\nabla_{(a}{\cal C}_{b)},
\label{e:ConstraintEqGauge}
\end{eqnarray}
which is also identical to the standard generalized harmonic case.  It
follows that the Pretorius-Gundlach~\cite{Pretorius2005c,
  Pretorius2005a, Gundlach2005} constraint-damping mechanism can be
applied to the covariant evolution system without modification.  In
particular, we add the constraint-damping terms:
\begin{eqnarray}
0=R_{ab}- \nabla_{(a}{\cal C}_{b)}
+\gamma_0\left[t_{(a}{\cal C}_{b)}-\half
\psi_{ab}t^c{\cal C}_c\right], 
\label{e:CovariantGH0}
\end{eqnarray}
where $t^a$ is a timelike vector field, and $\gamma_0$ is a constant.
The constraint evolution implied by the covariant evolution system
with constraint damping, Eq.~(\ref{e:CovariantGH0}), 
is obtained by using
the Bianchi identities.  The result is the evolution system
\begin{eqnarray}
0=\nabla^b\nabla_b{\cal C}_a -2\gamma_0\nabla^b\left[t_{(b}{\cal C}_{a)}\right]
+{\cal C}^b\nabla_{(a}{\cal C}_{b)}-\half\gamma_0 t_a
{\cal C}^b{\cal C}_b,\nonumber\\
\end{eqnarray}
which is a damped wave equation for small, short-wavelength ${\cal C}_a$ when
$\gamma_0>0$.
The covariant vacuum Einstein equation, including the 
constraint-damping terms, reduces therefore to the following manifestly
hyperbolic system:
\begin{eqnarray}
\psi^{cd}\tilde\nabla_c\tilde\nabla_d\psi_{ab}&=&
-2\nabla_{(a}H_{b)}-2\psi^{cd}\tilde R^{e}{}_{cd(a}\psi_{b)e}
\nonumber\\
&& + 2\psi^{cd}\psi^{ef}\left(\tilde\nabla_e\psi_{ca}
\tilde\nabla_f\psi_{bd}-\Delta_{ace}\Delta_{bdf}\right)
\nonumber\\
&& +\gamma_0\left[2\delta^c_{(a}t_{b)}-\psi_{ab}t^c\right]
\left(H_c+\Delta_c\right).
\label{e:CovariantGH}
\end{eqnarray}
This equation (minus the constraint-damping terms) was derived
previously by Ruiz, Rinne and Sarbach~\cite{Ruiz2007},
who used it in their analysis of boundary conditions, 
and by Brown~\cite{Brown2011}, who used it to derive an
action principle  for this 
second-order covariant generalized harmonic 
formulation of Einstein's equation.

The idea now is to transform Eq.~(\ref{e:CovariantGH}) into a
spatially covariant symmetric-hyperbolic first-order evolution system.
To that end, we introduce the physical timelike normal, $t^a$, which
satisfies $\psi_{ab}t^at^b=-1$, and which can be expressed in terms of the
lapse $N$ and shift $N^k$ of the physical metric: $t^a\partial_a =
N^{-1}(\partial_t - N^k\partial_k)$.  We then define the first-order
variables, $\Pi_{ab}$ and $\Phi_{iab}$:
\begin{eqnarray}
\Pi_{ab} &=& - t^c\tilde \nabla_c\psi_{ab},
\label{e:PiDef}\\
\Phi_{iab} &=& \tilde\nabla_i\psi_{ab},
\label{e:PhiDef}
\end{eqnarray}
where the indices $\scriptstyle\{i,j,k,...\}$ range only over the
spatial coordinates, while the indices $\scriptstyle \{a,b,c,d, ...
\}$ range over both space and time coordinates.  The introduction of
$\Phi_{iab}$ also implies the existence of a new constraint for the
system:
\begin{eqnarray}
{\cal C}_{iab}=\tilde\nabla_i\psi_{ab}-\Phi_{iab}.
\label{e:Constraint3index}
\end{eqnarray}
We note that the constraint, ${\cal C}_{iab}$, like the 
first-order evolution fields, $\Pi_{ab}$ and $\Phi_{iab}$, is a
tensor with respect to purely spatial coordinate transformations.

The spatially covariant first-order evolution equation for $\psi_{ab}$
follows directly from the definition of $\Pi_{ab}$ in
Eq.~(\ref{e:PiDef}):
\begin{eqnarray}
&&\partial_t\psi_{ab} - (1+\gamma_1)N^k\partial_k\psi_{ab}  \nonumber\\
&&\qquad =- N \Pi_{ab}-\gamma_1N^k\Phi_{kab}
-2(1+\gamma_1)N^k\tilde \Gamma{}^j_{k(a}\psi{}^{\vphantom{j}}_{b)j}.
\qquad
\label{e:CGHpsi}
\end{eqnarray}
The constraint term $\gamma_1 N^k{\cal C}_{kab}/N$, where $\gamma_1$
is an arbitrary constant, has been added to the definition of
$\Pi_{ab}$ to obtain Eq.~(\ref{e:CGHpsi}).  The particular
choice $\gamma_1=-1$ makes the system linearly degenerate,
which implies that shocks will not form from smooth
initial data~\cite{Liu1979}.
Here the quantity
$\tilde\Gamma^a_{bc}$ is the connection associated with the reference
metric $\tilde\psi_{ab}$.  We assume that this reference metric is
static, $\partial_t\tilde\psi_{ab}=0$, and that
$\tilde\psi_{tt}=-1$ and $\tilde\psi_{ti}=0$.
It follows that all of the time
components of $\tilde \Gamma^a_{bc}$ vanish, $\tilde\Gamma^t_{bc}
=\tilde\Gamma^a_{tc}=0$, in this case.

The spatially covariant first-order evolution equation for $\Pi_{ab}$
follows from the second-order covariant evolution equation,
Eq.~(\ref{e:CovariantGH}):
\begin{eqnarray}
&&\!\!\!\!\!\!\!\!\!
\partial_t\Pi_{ab} -N^k\partial_k\Pi_{ab} + N g^{ki}\partial_k\Phi_{iab}
-\gamma_1\gamma_2 N^k\partial_k\psi_{ab}
\nonumber\\
&&
= 2N\psi^{cd}\left(g^{ij}\Phi_{ica}\Phi_{jdb}-\Pi_{ca}\Pi_{db}
-\psi^{ef}\Delta_{ace}\Delta_{bdf}\right)
\nonumber\\
&&\quad
-2N\nabla_{(a}H_{b)}-\half N t^c t^d\Pi_{cd}\Pi_{ab}
-N t^c\Pi_{ci}g^{ij}\Phi_{jab}
\nonumber\\
&&\quad
+N\gamma_0\left[2\delta^{c}_{(a}t^{\vphantom{c}}_{b)}
-\psi_{ab}t^c\right]\left(H_c+\Delta_c\right)
-\gamma_1\gamma_2N^i\Phi_{iab}
\nonumber\\
&&\quad 
-2N\psi^{ij}\tilde R^k{}_{ij(a}\psi_{b)k} 
-2N^i\tilde \Gamma^j_{i(a}\Pi^{\vphantom{j}}_{b)j}
+N g^{ij}\tilde \Gamma^k_{ij}\Phi_{kab}
\nonumber\\
&&\quad 
+2N g^{ij}\Phi_{ik(a}\tilde\Gamma^k_{b)j}
-2\gamma_1\gamma_2N^i\tilde\Gamma^j_{i(a}\psi^{\vphantom{j}}_{b)j}
\nonumber\\
&&\quad -8\pi N(2 T_{ab}-\psi_{ab}\psi^{cd}T_{cd}) - 2 N \Lambda \psi_{ab}
.
\label{e:CGHPi}
\end{eqnarray}
In this expression, $T_{ab}$ represents the stress-energy tensor of any
matter that may be present in the solution, and $\Lambda$ is the
cosmological constant.  We use the notation $g_{ab}$ for the spatial
metric, $g_{ab}=\psi_{ab}+t_at_b$, which satisfies $g_{ab}t^b=0$.  The
quantity $g^{ij}$ is the inverse of the spatial metric
$g_{ij}=\psi_{ij}$.  The quantities $\Delta_{abc}$ and $\Delta_a=
\psi^{bc}\Delta_{abc}$ that appear on the right side of
Eq.~(\ref{e:CGHPi}) are to be written as functions of the first-order
fields $\Pi_{ab}$ and $\Phi_{iab}$: i.e., the derivatives
$\tilde\nabla_a\psi_{bc}$ that appear in the definition of
$\Delta_{abc}$, Eq.~(\ref{e:DeltaDef}), are to be replaced by the
expressions
\begin{eqnarray}
\tilde\nabla_t\psi_{ab} &=& -N\Pi_{ab} + N^i\Phi_{iab},\\
\tilde\nabla_i\psi_{ab} &=& \Phi_{iab}.
\end{eqnarray}
The derivation of the evolution equation for $\Pi_{ab}$, 
Eq.~(\ref{e:CGHPi}), also uses the identity $t^b\tilde\nabla_b t^a
=\half t^c(2\psi^{ab}+t^at^b)\Pi_{bc}$.

The spatially covariant first-order evolution equation for
$\Phi_{iab}$ is obtained by requiring that the constraint 
${\cal C}_{iab}$ satisfy a damped, advection-type evolution equation:
\begin{eqnarray}
t^c\tilde\nabla_c{\cal C}_{iab} = - \gamma_2 {\cal C}_{iab}.
\label{e:ConstraintEq3index}
\end{eqnarray}
Choosing the constant $\gamma_2>0$ ensures 
that the constraint 
${\cal C}_{iab}$ is driven toward zero as the system evolves.  This
constraint-damping equation implies the following first-order
evolution equation for $\Phi_{iab}$:
\begin{eqnarray}
&&\!\!\!\!\!\!\!\!\!
\partial_t\Phi_{iab}-N^k\partial_k\Phi_{iab}+N\partial_i\Pi_{ab}
-N\gamma_2\partial_i\psi_{ab}
\nonumber\\
&&=\half N t^c t^d \Phi_{icd}\Pi_{ab} 
+ N g^{jk}t^c\Phi_{ijc}\Phi_{kab}-N\gamma_2 \Phi_{iab}
\nonumber\\
&&\quad-N^j\tilde\Gamma^k_{ij}\Phi_{kab}
-2N^j\Phi_{ik(a}\tilde\Gamma^k_{b)j}
+2N\tilde\Gamma^{j}_{i(a}\Pi^{\vphantom{j}}_{b)j}
\nonumber\\
&&\quad-2N\gamma_2\tilde\Gamma^j_{i(a}\psi^{\vphantom{j}}_{b)j}
-2N^k\psi_{j(a}\tilde R^j{}_{b)ik}.
\label{e:CGHPhi}
\end{eqnarray}
The derivation of this evolution equation uses the identity
$\tilde\nabla_it^a = -\half t^c(2\psi^{ab}+t^at^b)\,\Phi_{ibc}$.

The principal parts of a first-order evolution system are defined to
be the terms that involve the derivatives of the fields.  We use the
notation 
$\partial_t u^{\cal A} + 
A^{k{\cal A}}{}_{\cal B}(\mathbf{x},\mathbf{u})\,
\tilde\nabla_k u^{\cal B} \simeq 0$ to denote
the principal parts of the general first-order hyperbolic system 
described in
Eq.~(\ref{e:FirstOrderHyperbolicEq}).  The principal parts of the
spatially covariant first-order evolution system defined in
Eqs.~(\ref{e:CGHpsi}), (\ref{e:CGHPi}), and (\ref{e:CGHPhi}) are
therefore given by
\begin{eqnarray}
&&
\!\!\!\!\!\!\!\!
\partial_t\psi_{ab} - (1+\gamma_1)N^k\tilde\nabla_k\psi_{ab}  \simeq 0,
\nonumber\\
&&
\!\!\!\!\!\!\!\!
\partial_t\Pi_{ab} -N^k\tilde\nabla_k\Pi_{ab} + N g^{ki}\tilde\nabla_k\Phi_{iab}
-\gamma_1\gamma_2 N^k\tilde\nabla_k\psi_{ab}\simeq 0,
\nonumber\\
&&
\!\!\!\!\!\!\!\!
\partial_t\Phi_{iab}-N^k\tilde\nabla_k\Phi_{iab}+N\tilde\nabla_i\Pi_{ab}
-N\gamma_2\tilde\nabla_i\psi_{ab}\simeq 0.\nonumber
\end{eqnarray}
These terms are identical to the principal parts of the standard
first-order generalized harmonic evolution system described in
Ref.~\cite{Lindblom2006}.  It follows that this spatially covariant
first-order evolution system is symmetric hyperbolic with the standard
symmetrizer~\cite{Lindblom2006}:
\begin{eqnarray}
S_{\alpha\beta} du^\alpha du^\beta &=&  m^{ab}m^{cd}\bigl(
L^{-2} d\psi_{ac}d\psi_{bd}
+d\Pi_{ac}d\Pi_{bd}\nonumber\\
&&
-2\gamma_2d\psi_{ac}d\Pi_{bd}
+g^{ij}d\Phi_{iac}d\Phi_{jbd}\bigr),
\end{eqnarray}
where $m^{ab}$ is any positive-definite metric (e.g.,
$m^{ab}=g^{ab}+t^a t^b$, or even $m^{ab}=\delta^{ab}$) and $L$ is a
constant with the dimension of a length.  It follows that the
characteristic fields and  speeds of the
spatially covariant first-order evolution system 
are identical to those of the noncovariant generalized harmonic
system.  In particular, the
characteristic fields $\hat u^{\cal K}=\{\hat u^0_{ab},\hat u^{1\pm}_{ab},
\hat u^2_{iab}\}$ are given by
\begin{eqnarray}
\hat u^{0}_{ab}      &=& \psi_{ab},\\
\hat u^{{1}\pm}_{ab} &=& \Pi_{ab} \pm n^i \Phi_{iab}
                           -\gamma_2 \psi_{ab}, \\
\hat u^{2}_{iab}     &=& P_i{}^k \Phi_{kab},
\end{eqnarray}
\noindent
where $P_i{}^k=\delta_i{}^k - n_i n^k$.  All of these 
characteristic fields
are tensors with respect to spatial 
coordinate transformations.
The characteristic fields $\hat u^{0}_{ab}$ have coordinate
characteristic speed $-(1+\gamma_1)n_kN^k$, the fields $\hat
u^{{1}\pm}_{ab}$ have speeds $-n_kN^k\pm N$, and the fields $\hat
u^{2}_{iab}$ have speed $-n_kN^k$.

The first-order dynamical fields $\Pi_{ab}$ and $\Phi_{iab}$ of the
spatially covariant first-order evolution system are different from
those used in the noncovariant generalized-harmonic evolution
equations.  These differences require that additional terms
proportional to the reference connection $\tilde \Gamma^a_{bc}$ and
its curvature $\tilde R^a_{bcd}$ be added to the 
right sides of 
Eqs.~(\ref{e:CGHpsi}), (\ref{e:CGHPi}), and
(\ref{e:CGHPhi}).  
But these additional terms do not affect the
principal parts of the equations, the expressions for the
characteristic fields in terms of the dynamical fields, or the
characteristic speeds of the system.  We also note that the reference
metric can be chosen to be the Minkowski metric,
$\tilde\psi_{ab}=\eta_{ab}$, on manifolds that admit a global flat
metric (e.g., manifolds 
whose spatial slices are subsets 
of $R^3$).  When
expressed in terms of the global Cartesian coordinates that are
available in such a case, the reference connection 
$\tilde\Gamma^a_{bc}$
and the reference curvature $\tilde R^a{}_{bcd}$ both vanish
identically.  
The spatially covariant first-order
evolution system is then precisely the same as the 
standard noncovariant
generalized harmonic system.  The standard first-order generalized
harmonic system is therefore 
a special case of the new covariant
first-order system on manifolds that admit a flat reference metric.

The constraints ${\cal C}_a$ and ${\cal C}_{iab}$ defined in
Eqs.~(\ref{e:ConstraintGauge}) and (\ref{e:Constraint3index}) evolve
according to Eqs.~(\ref{e:ConstraintEqGauge}) and
(\ref{e:ConstraintEq3index}).  As in the noncovariant generalized
harmonic evolution system~\cite{Lindblom2006}, the second-order
evolution system for these constraints can be converted into a
symmetric-hyperbolic first-order system by adding the following
secondary constraints:
\begin{eqnarray}
{\cal F}_a &=& t^c\nabla_c{\cal C}_a,
\label{e:FConstraintDef}\\
{\cal C}_{ia}&=& \nabla_i{\cal C}_a,
\label{e:TwoIndexConstraintDef}\\
{\cal C}_{ijab}&=& 2\tilde\nabla_{[i}{\cal C}_{j]ab}.
\label{e:FourIndexConstraintDef}
\end{eqnarray}
Expressions for all the constraints $\mathcal{C}_a$,
$\mathcal{C}_{iab}$, $\mathcal{F}_a$, $\mathcal{C}_{ia}$, and
$\mathcal{C}_{ijab}$ are given in Appendix~\ref{s:AppendixA} in terms
of the dynamical fields of the system $u^{\cal
  A}=\{\psi_{ab},\Pi_{ab},\Phi_{iab}\}$ and their spatial derivatives.

\section{Einstein-Klein-Gordon Static Universe}
\label{s:EinsteinStatic}

The remainder of this paper is devoted to performing a number of
simple numerical tests on the multicube methods described in
Sec.~\ref{s:ReviewMultiCubeMethod}, using the spatially
covariant representation of the Einstein system developed in
Sec.~\ref{s:CovariantEinsteinSystem}. Our primary goal here is to verify
that our implementation of these methods in the SpEC code (developed
by the SXS Collaboration, originally at Caltech and 
Cornell~\cite{Kidder2000a, Scheel2006, Scheel2009,
  Szilagyi:2009qz}) is numerically stable and convergent for 
long-time-scale evolutions.  
Most known solutions to Einstein's equation on manifolds
with compact spatial topologies collapse to a singularity or expand
exponentially without bound on very short time scales.  
Neither of
these types of solutions is well suited for testing the long-term
stability of a numerical code.  We have therefore focused our
attention on one of the few known
time-independent solutions on a manifold
with compact spatial topology: the Einstein static universe.

The Einstein static universe is a time-independent (static) and
spatially homogeneous solution to Einstein's equation on the manifold
$R \times S^3$:
\begin{eqnarray}
ds^2 &=& \psi^0_{ab}dx^a dx^b \nonumber\\
&\equiv & -dt^2 + R^2_3\left[d\chi^2
+ \sin^2\chi \left(d\theta^2 + \sin^2\theta d\varphi^2\right)\right].
\qquad\label{e:EinsteinStaticMetric}
\end{eqnarray}
The spatial part of this geometry is just the standard 
round metric on $S^3$.  This metric satisfies Einstein's gravitational field
equation with source
\begin{eqnarray}
R_{ab} -\half\psi_{ab}R +\Lambda \psi_{ab} = 8\pi T_{ab},
\end{eqnarray}
where $\Lambda$ is the 
cosmological constant and $T_{ab}$ is the
stress-energy tensor of the matter present in the spacetime.  
The cosmological constant has the value $\Lambda=1/R_3^2$ for the Einstein
static universe, while the stress-energy tensor $T_{ab}=\rho\,
\partial_a t \partial_b t$ corresponds to a pressureless ``dust''
with $\rho=1/4\pi R_3^2$.  Dynamical evolutions of spacetimes
containing dust typically develop shell-crossing
singularities~\cite{Tolman1934}. 
Hence, dust is not particularly well suited for numerical tests 
using spectral methods, which require smooth solutions
to achieve exponential convergence~\cite{Boyd1999}. 

An alternate interpretation of the Einstein static universe can be
constructed in which the matter part of the solution is generated by a
complex Klein-Gordon scalar field instead of dust.  The stress-energy
tensor of a complex scalar field $\phi$ is given by
\begin{eqnarray} 
T_{ab}&=&\half\left(\nabla_a\phi \nabla_b\phi^*
+\nabla_a\phi^* \nabla_b\phi\right)
\nonumber\\
&&-\half\psi_{ab}\left(\psi^{cd}\nabla_c\phi\nabla_d\phi^*
+\mu^2\phi\,\phi^*\right),
\end{eqnarray}
where $\phi^*$ is the complex conjugate of the field, and $\mu$ is
its mass.  This field satisfies the covariant Klein-Gordon equation,
\begin{eqnarray}
\nabla^a\nabla_a\phi = \mu^2\phi,
\end{eqnarray}
as a consequence of the stress-energy conservation law
$\nabla^aT_{ab}=0$.  One solution to this scalar field equation in the
Einstein static universe is 
\begin{eqnarray}
\phi=\phi_0\,e^{i\mu t}, \label{e:phi0reference}
\end{eqnarray} 
where $\phi_0$ is a (complex) constant.  This particular solution has
a stress-energy tensor that can be used as the source term needed for
an Einstein-Klein-Gordon static universe by taking $\Lambda=1/R_3^2$
and $\mu^2|\phi_0|^2 = 1/4\pi R_3^2$.  Note that only the product
$|\phi_0|\mu$ is fixed, not their individual values.  For our
numerical tests, we use $\mu=2/R_3$ so that $|\phi_0| =1/\sqrt{16\pi}$.
Also note that although the geometry of the Einstein-Klein-Gordon
universe is static, the scalar field $\phi$ oscillates with frequency
$\mu$.   In our numerical test evolutions, we use the
value $R_3=1$ for the scale of the $S^3$ geometry.

The first test of our implementation of the multicube methods
described in Sec.~\ref{s:ReviewMultiCubeMethod} is to evolve initial
data for the coupled Einstein and Klein-Gordon evolution equations
based on the static Einstein-Klein-Gordon universe solution.  The
spacetime manifold for this solution has the topology $R\times S^3$,
so we use the round metric $\psi^0_{ab}$ of
Eq.~(\ref{e:EinsteinStaticMetric}) as our smooth reference metric:
$\tilde\psi_{ab}=\psi^0_{ab}$. The initial data for the dynamical
fields of the Einstein evolution system,
$u^\alpha=\{\psi_{ab},\Pi_{ab},\Phi_{iab}\}$, are constructed from the
metric of the Einstein static universe solution.  In particular, we
take $\psi_{ab}=\psi^0_{ab}$ and $\Pi_{ab}=\Phi_{iab}=0$ 
initially.
The dynamical fields of the complex first-order Klein-Gordon system
consist of the fields $u^\alpha_\phi=\{\phi,\Pi^\phi,\Phi^\phi_i\}$.
The initial values of these fields for the Einstein-Klein-Gordon
static universe solution are given by $\phi=\phi_0$,
$\Pi^\phi=-i\mu\phi_0$, and $\Phi^\phi_i=0$.  
We carry out the numerical 
evolutions of these fields using the
multicube representation of $S^3$ developed in 
Ref.~\cite{LindblomSzilagyi2011a}, which gives the explicit multicube
expressions for the metric $\psi^0_{ab}$, as well as 
the standard three-sphere angular coordinates
$\chi$, $\theta$, and $\varphi$, in terms of the global multicube
Cartesian coordinates.

Evolutions of Einstein's equation require 
appropriate gauge (i.e., coordinate) conditions to be specified.  
The gauge is
specified in the spatially covariant first-order representation of the
Einstein equation, described in Sec.~\ref{s:CovariantEinsteinSystem},
using the gauge source covector $H_a$.
The gauge condition is imposed
with the covariant generalized harmonic condition: $H_a =
-\Delta_{abc}\psi^{bc}$.  It is straightforward to show that the
static Einstein-Klein-Gordon solution satisfies this condition with
$H_a=0$.  The gauge choices used in our numerical tests are harmonic
gauge for the time coordinate and damped harmonic
gauge~\cite{Lindblom2009c} for the spatial coordinates:
\begin{eqnarray}
H_t &=& 0,\\
H_i &=& - \mu_G N_i / N,
\label{e:DampendHarmonicGauge}
\end{eqnarray}
where $\mu_G$ is a constant that serves as the harmonic gauge damping
parameter, $N$ is the lapse, and $N_i$ is the shift of the spacetime
metric.  This choice of gauge source function $H_a$ depends only on
the spacetime metric (and not its derivatives), 
so the covariant first-order representation of
Einstein's equation is hyperbolic in this case.
Note that this choice of gauge
reduces to harmonic gauge $H_a=0$ for the Einstein-Klein-Gordon static
universe solution where $N=1$ and $N_i=0$.

\begin{figure}
\includegraphics[width=3.4in]{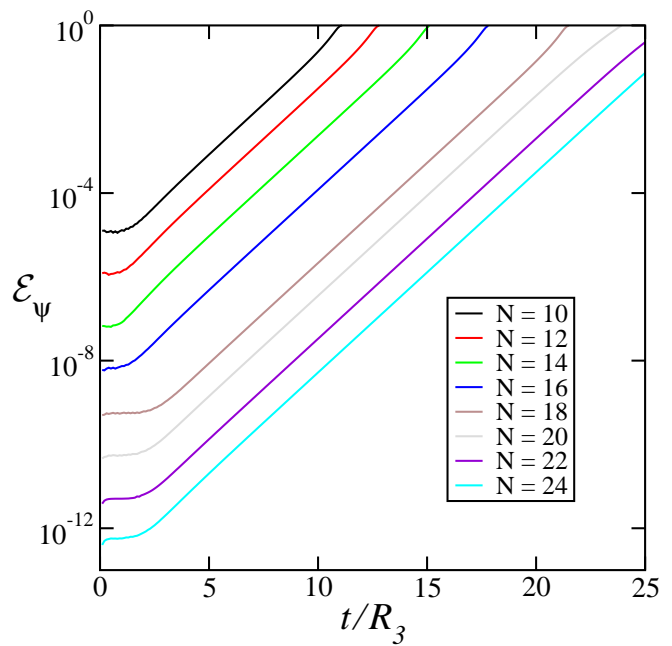}
\caption{\label{f:EinsteinStaticNoDampErrPsi} Errors in the numerical
  evolution of the metric $\psi_{ab}$ using initial data for the
  Einstein-Klein-Gordon static solution.  Numerical resolution
used in each spatial dimension of each cubic region is denoted by $N$.}
\end{figure}

\begin{figure}
\includegraphics[width=3.4in]{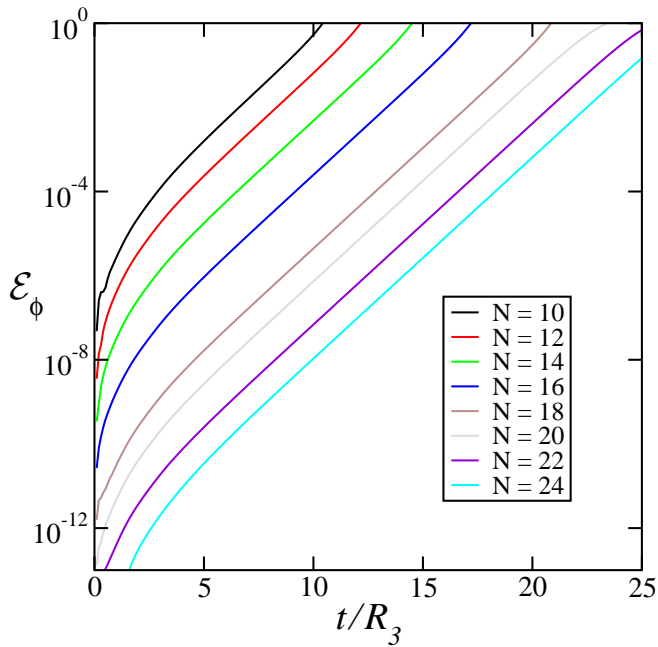}
\caption{\label{f:EinsteinStaticNoDampErrSwPsi} Errors in the
  numerical evolution of the complex Klein-Gordon scalar field $\phi$
  using initial data for the Einstein-Klein-Gordon static solution.
 Numerical resolution
 used in each spatial dimension of each cubic region is denoted by $N$.}
\end{figure}

\begin{figure}
\includegraphics[width=3.4in]{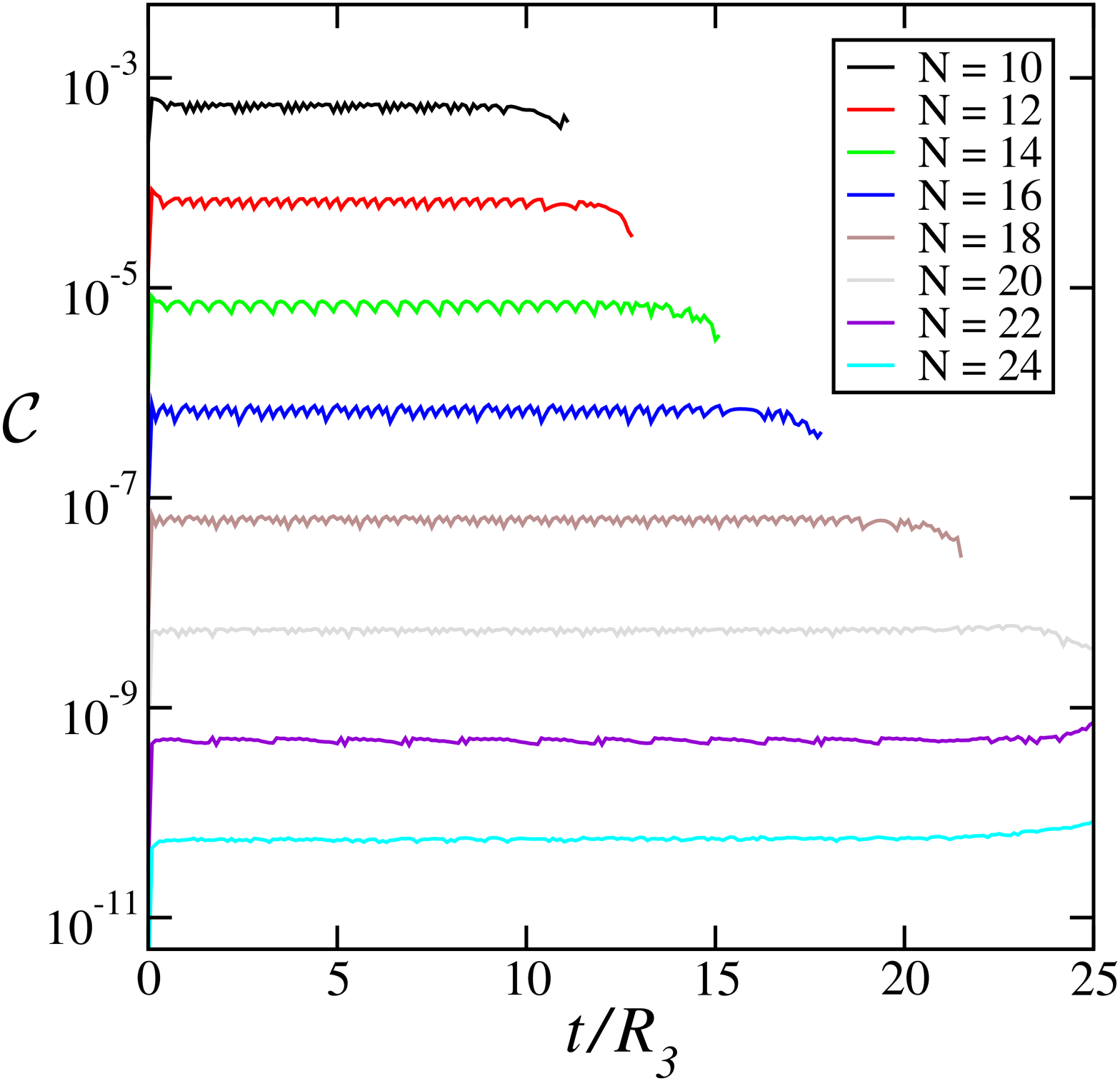}
\caption{\label{f:EinsteinStaticNoDampGhCe} 
  Constraint norm
  ${\cal C}$ in
  the numerical evolutions using initial data for the
  Einstein-Klein-Gordon static solution.  Numerical resolution
used in each spatial dimension of each cubic region is denoted by $N$.}
\end{figure}

The results of this first numerical test are illustrated in
Figs.~\ref{f:EinsteinStaticNoDampErrPsi}--\ref{f:EinsteinStaticNoDampGhCe}.
Figure~\ref{f:EinsteinStaticNoDampErrPsi} shows the error in the
metric $\mathcal{E}_\psi$ as a function of time for evolutions using
different spatial resolutions.  The constant $N$, which
appears in the labels of these figures, is the number of spectral
basis functions used in the solution for each dimension of each cubic
region ${\cal B}_A$.  The error measure $\mathcal{E}_\psi$ is defined
by
\begin{equation}
  \mathcal{E}_\psi^2 \equiv 
  \frac{\int m^{ab}m^{cd}\Delta\psi_{ac}\Delta\psi_{bd} \sqrt{g}\, d^{\,3}x}
       {\int m^{ab}m^{cd}\psi_{ac}^\mathcal{N}\psi_{bd}^\mathcal{N} 
\sqrt{g}\, d^{\,3}x},
\label{e:EpsiDef} 
\end{equation}
where $\Delta\psi_{ab}=\psi_{ac}^\mathcal{N}-\psi_{ac}^\mathcal{A}$,
$\psi_{ab}^\mathcal{A}$, and $\psi_{ab}^\mathcal{N}$ represent the
analytically and numerically determined metrics, and $m^{ab}$ is a
positive definite tensor, taken here to be $m^{ab}=\delta^{ab}$ in the
global multicube Cartesian coordinates.  This quantity measures the
fractional accuracy of the numerically determined metric.  Similarly,
Fig.~\ref{f:EinsteinStaticNoDampErrSwPsi} shows the scalar field error
measure, $\mathcal{E}_\phi$, defined by
\begin{equation}
  \mathcal{E}_\phi^2 \equiv 
  \frac{\int |\Delta\phi|^2 \sqrt{g}\, d^{\,3}x}
       {\int |\phi^\mathcal{N}|^2 \sqrt{g}\, d^{\,3}x},
\label{e:EphiDef}
\end{equation}
where $\Delta\phi=\phi^\mathcal{N}-\phi^\mathcal{A}$, and vertical
bars denote the complex absolute value.
Figure~\ref{f:EinsteinStaticNoDampGhCe} shows the constraint errors of
the combined Einstein and Klein-Gordon evolution equations.  We
combine these constraint errors into the single quantity $\mathcal{C}$,
defined by
\begin{equation}
  \mathcal{C}^2 \equiv 
  \frac{\int \mathcal{C}_\psi^2 \sqrt{g}\, d^{\,3}x}
       {\int \mathcal{N}_\psi^{\,2} \sqrt{g}\, d^{\,3}x}
       +\frac{\int \mathcal{C}_\phi^2 \sqrt{g}\, d^{\,3}x}
       {\int \mathcal{N}_\phi^{\,2} \sqrt{g}\, d^{\,3}x}.
\end{equation}
The quantity $\mathcal{C}_\psi$ measures the size of the constraint
violations of the Einstein system, and $\mathcal{N}_\psi$ measures the
sizes of the spatial derivatives of the dynamical fields:
\begin{align}
  \begin{split}
    \mathcal{C}^2_\psi &\equiv m^{ab}\Big( \mathcal{C}_a\mathcal{C}_b
    + \mathcal{F}_a\mathcal{F}_b 
    + \tilde g^{ij} m^{cd} \big[ \mathcal{C}_{iac}\mathcal{C}_{jbd} \\
    & \qquad + \fourth \tilde g^{kl} \mathcal{C}_{ikac} 
      \mathcal{C}_{jlbd} \big]
    \Big),
  \end{split} \\
  \begin{split}
    \mathcal{N}^{\,2}_\psi &\equiv m^{ab} m^{cd} \tilde g^{ij} \Big( 
    \partial_i \psi_{ac} \partial_j \psi_{bd}
    + \partial_i \Pi_{ac} \partial_j \Pi_{bd} \\
    & \qquad + \tilde g^{kl} \partial_i \Phi_{kac} \partial_j \Phi_{lbd}
    \Big).
  \end{split}
\end{align}
The constraints of the Einstein evolution system used 
to construct $\mathcal{C}_\psi$ are defined in Eqs.~(\ref{e:ConstraintGauge}),
(\ref{e:Constraint3index}), (\ref{e:FConstraintDef}),
(\ref{e:TwoIndexConstraintDef}), and (\ref{e:FourIndexConstraintDef}).
The dimensionless ratio between the norms of $\mathcal{C}_\psi$ and
$\mathcal{N}_\psi$ is designed to give a meaningful measure of the
fractional errors due to  
constraint violations of the Einstein system.
The quantities $\mathcal{C}_\phi$ and $\mathcal{N}_\phi$, defined by
\begin{align}
  \mathcal{C}^2_\phi &\equiv m^{ij} \Big( \mathcal{C}^\phi_i
  \mathcal{C}^\phi_j + \half m^{kl} \mathcal{C}^\phi_{ik}
  \mathcal{C}^\phi_{jl} \Big), \\ \mathcal{N}^{\,2}_\phi &\equiv \mu^2
  |\phi|^2,
\label{e:CErrorDef}
\end{align}
play analogous roles for the Klein-Gordon evolution system.
The scalar field constraints $\mathcal{C}^\phi_i$ and
$\mathcal{C}^\phi_{ij}$ used to construct $\mathcal{C}_\phi$ are
defined by $\mathcal{C}^\phi_i=\Phi^\phi_i-\tilde \nabla_i\phi$ and
$\mathcal{C}^\phi_{ij}=\tilde\nabla_i\Phi^\phi_j-\tilde\nabla_j\Phi^\phi_i$.

Figures~\ref{f:EinsteinStaticNoDampErrPsi} and
\ref{f:EinsteinStaticNoDampErrSwPsi} show that our numerical solutions
diverge exponentially away from the Einstein-Klein-Gordon static
universe solution, while Fig.~\ref{f:EinsteinStaticNoDampGhCe} shows
that the constraints are well satisfied 
during a time in which 
this instability grows by over 10 orders of
magnitude.
Our numerical evolutions therefore 
confirm the existence of the
instability of the Einstein static universe first noted by
Eddington~\cite{Eddington1030}.  The growth rate of this instability
can be measured numerically from our evolutions, giving
$1/\tau_\mathcal{N}\approx1.100501(1)$, where the number in 
parentheses represents the estimated uncertainty in the last digit.
This agrees with the analytical value,
$1/\tau_\mathcal{A}=\sqrt{2\sqrt{13}-6}\approx 1.1005010$, computed
for this unstable mode in Sec.~\ref{s:PerturbedEinsteinStatic}.

\section{Mode Damping}
\label{s:ModeDamping}

The straightforward numerical test of the Einstein-Klein-Gordon
evolution system described in Sec.~\ref{s:EinsteinStatic} confirms
that our implementation of the multicube method is basically correct
and that our numerical methods are basically stable and convergent.  
Unfortunately,
those evolutions persist for just a few light-crossing times of the
$S^3$ geometry.  These first tests do not, therefore, allow us to
identify more subtle errors that might become evident only on much longer
time scales.  Nor do they test our implementation on solutions having
more complicated spatial and temporal structures than the spatially
homogeneous Einstein-Klein-Gordon static universe.  We address these
shortcomings in the following sections by performing more challenging
variations on our original Einstein-Klein-Gordon static universe test.

In this section we construct small, unphysical 
damping forces that
suppress the growth of the modes responsible for the Eddington
instability.  
The modified evolution equations can be written abstractly
in the form
\begin{eqnarray}
\partial_t\psi_{ab} &=& f_{ab}+\mathcal{D} f_{ab},\label{e:psidot}\\
\partial_t\Pi_{ab} &=& F_{ab}+\mathcal{D} F_{ab},\label{e:Pidot}\\
\partial_t\phi &=& f_\phi+\mathcal{D} f_\phi,\label{e:phidot}\\
\partial_t\Pi_\varphi &=& F_\phi+\mathcal{D} F_\phi,\label{e:Phiphidot}
\end{eqnarray}
where $f_{ab}$, $F_{ab}$, $f_\varphi$ and $F_\varphi$ are the
expressions for the right sides of the unmodified Einstein-Klein-Gordon 
evolution equations, while  $\mathcal{D}f_{ab}$,
$\mathcal{D}F_{ab}$, $\mathcal{D}f_\varphi$ and
$\mathcal{D}F_\varphi$ represent the unphysical mode-damping forces.

Any physical mode, in particular the one responsible for the Eddington
instability, has a certain very specific spatial structure.  This fact
is used in this section to construct mode-damping forces
that suppress the degrees of freedom of the system having that
particular structure, while leaving unaffected the other dynamical
degrees of freedom of the system.  The effectiveness of the resulting
mode-damping forces is then tested by evolving initial data for the
Einstein-Klein-Gordon static universe solution. These tests confirm
the effectiveness of these mode-damping forces.  More importantly,
these tests also confirm the numerical stability and convergence of
our implementation of the multicube method for solving Einstein's
equation over very long time scales.

The most convenient and efficient way to represent the spatial
structures of tensor fields on $S^3$ is to expand those fields in the
tensor harmonics of the three-sphere~\cite{Sandberg1978}.  The basic
properties of the scalar, vector, and rank-2 tensor three-sphere
harmonics that are relevant to our work here are summarized in
Appendix~\ref{s:AppendixB}. The particular harmonics
that play an important role in the unstable modes of the
Einstein-Klein-Gordon static universe are the scalar harmonics
$Y^{k\ell m}$ and the vector harmonics $\tilde \nabla_i Y^{k\ell m}$.
The time-dependent projections of a scalar field $Q(\vec x, t)$ and a
vector field $V_i(\vec x, t)$ onto these harmonics are defined, 
respectively, as
\begin{eqnarray}
Q^{k\ell m}(t) &=& \int Q(\vec x,t) Y^{*k\ell m}\sqrt{\tilde g}\,d^{\,3}x,
\label{e:ScalarProjection}\\
V^{k\ell m}(t) &=& \int \tilde g^{ij}V_i(\vec x,t) \tilde \nabla_j Y^{*k\ell m} 
\sqrt{\tilde g}\,d^{\,3}x,\label{e:VectorProjection}
\end{eqnarray}
where $Y^{*k\ell m}$ in these equations denotes the complex conjugate.

The mode responsible for the Eddington instability is spatially
homogeneous, like the Einstein-Klein-Gordon solution itself.
Therefore, the spatial structures 
of the dynamical fields for this mode
are completely described by 
the $k=\ell=m=0$ three-sphere harmonics.
The mode-damping forces needed to suppress the growth of this
instability can therefore be constructed using only the $k=\ell=m=0$
three-sphere harmonic projections of the quantities $\psi=\tilde
g^{ij}\psi_{ij}$, $f=\tilde g^{ij} f_{ij}$, $\psi_{tt}$, $f_{tt}$,
$\Pi=\tilde g^{ij}\Pi_{ij}$, $F=\tilde g^{ij}F_{ij}$, $\Pi_{tt}$,
$F_{tt}$, $\phi$, $f_\phi$, $\Pi_\phi$, and $F_\phi$.  We use these
three-sphere harmonic projections to construct the following
mode-damping forces:
\begin{eqnarray}
\mathcal{D} f_{ab}^{\,000}&\!\!\!\!\equiv& \!\!\!\!- \frac{Y^{\,000}}{3R_3^3}
\left\{f^{\,000}(t) +
  \eta_G [\psi^{\,000}(t) - \psi^{\,000}(0)]\right\}\tilde g_{ab}
\nonumber\\ 
&&
\!\!\!\!\!\!\!\!\!\!\!\!\!\!\!\!\!\!\!
-\frac{Y^{\,000}}{R_3^3}\left\{f_{tt}^{\,000}(t) 
+\eta_G[\psi_{tt}^{\,000}(t)-\psi_{tt}^{\,000}(0)]\right\}\hat t_a \hat t_b, 
\quad\label{e:psi000Damping}
\\ 
\mathcal{D}F_{ab}^{\,000}&\equiv&
- \frac{Y^{\,000}}{3R_3^3}
\left[F^{\,000}(t) + \eta_G \,\Pi^{\,000}(t) \right]\tilde g_{ab}
\nonumber\\
&&
-\frac{Y^{\,000}}{R_3^3}
\left[F_{tt}^{\,000}(t) +\eta_G \,\Pi_{tt}^{\,000}(t)\right]\hat t_a \hat t_b,
\label{e:Pi000Damping}
\\
\mathcal{D}f_\phi^{\,000} &\equiv& - \frac{Y^{\,000}}{R_3^3}
\left\{\left[f_\phi^{\,000}(t) -i\mu \phi^{\,000}(0)e^{i\mu t}\right]\right.
\nonumber\\
&&\qquad\,\,\,
+ \left.\eta_S\left[\phi^{\,000}(t)-\phi^{\,000}(0)e^{i\mu t}\right]\right\},
\label{e:phi000Damping}
\end{eqnarray}
\begin{eqnarray}
\mathcal{D} F_\phi^{\,000} &\equiv& -\frac{Y^{\,000}}{R_3^3}
\left\{\left[F_\phi^{\,000}(t)
-i\mu \Pi_\phi^{\,000}(0)e^{i\mu t}\right] \right.
\nonumber\\
&&\quad\,\,\,
+\left.\eta_S\left[\Pi_\phi^{\,000}(t) 
-\Pi_\phi^{\,000} (0)e^{i\mu t}\right]\right\},
\label{e:Piphi000Damping}
\end{eqnarray}
where $\hat t_a=\partial_a t$.  The constants $\eta_G$ and $\eta_S$ in
these equations are damping rates (of order unity) that control how
quickly the mode damping acts to drive the $k=\ell=m=0$ component of
these solutions back toward their equilibrium values.  

It is straightforward to show that the modified Einstein-Klein-Gordon
evolution equations suppress the dynamics of the $k=\ell=m=0$ degrees
of freedom of the system, without affecting 
the dynamics in any other mode.  
Multiplying Eqs.~(\ref{e:psidot})--(\ref{e:Phiphidot}) by
$Y^{*000}$ and integrating the scalar parts (i.e., the spatial trace
and the $\scriptstyle tt$ components) over the $S^3$ geometry results
in the following equations for the $k=\ell=m=0$ components of the
various dynamical fields:
\begin{eqnarray}
&&\partial_t\left[\psi^{\,000}(t)-\psi^{\,000}(0)\right] =\nonumber\\
&&\qquad\qquad\qquad \qquad
 -\eta_G\left[\psi^{\,000}(t)-\psi^{\,000}(0)\right],
\label{e:K0psiDamping}\\ 
&&\partial_t\left[\psi^{\,000}_{tt}(t)-\psi^{\,000}_{tt}(0)\right]  =\nonumber\\ 
&&\qquad\qquad\qquad  \qquad
-\eta_G\left[\psi^{\,000}_{tt}(t)-\psi^{\,000}_{tt}(0)\right],\\ 
&&\partial_t\Pi^{\,000}(t)= -\eta_G\Pi^{\,000}(t),\\ 
&&\partial_t\Pi^{\,000}_{tt}(t) =-\eta_G\Pi^{\,000}_{tt}(t),\\ 
&&\partial_t\left[\phi^{\,000}(t)-\phi^{\,000}(0)e^{i\mu t}\right] =
\nonumber\\
&&\qquad\qquad\qquad \qquad
-\eta_S \left[\phi^{\,000}(t)-\phi^{\,000}(0)e^{i\mu t}\right],\\ 
&&\partial_t\left[\Pi_\varphi^{\,000}(t)-\Pi_\varphi^{\,000}(0)e^{i\mu t}\right] 
=\nonumber\\
&&\qquad\qquad\qquad \qquad
-\eta_S
\left[\Pi_\varphi^{\,000}(t)-\Pi_\varphi^{\,000}(0)e^{i\mu t}\right].\qquad
\label{e:K0PiphiDamping}
\end{eqnarray}
These equations drive the $k=\ell=m=0$ components of the various
dynamical fields toward their initial values.

Initial data for the Klein-Gordon static universe solution have been
evolved with the modified equations that include the $k=\ell=m=0$ 
mode-damping forces defined in
Eqs.~(\ref{e:psi000Damping})--(\ref{e:Piphi000Damping}).
Unfortunately, the resulting evolutions are still unstable.  The
numerically determined growth rate of this new instability is
$1/\tau_\mathcal{N}\approx 0.6180(1)$, where the number in parentheses
represents the estimated uncertainty in the last digit.  This agrees
with the analytical value, $1/\tau_\mathcal{A} =
(\sqrt{4+\mu_G^2R_3^2}-\mu_GR_3)/2 = (\sqrt{5}-1)/2\approx 0.618034$,
computed for an unstable $k=1$ mode of this system in
Sec.~\ref{s:PerturbedEinsteinStatic}.  The growth rate of this new
unstable mode is set by the constant $\mu_G$ (taken to have the value
$\mu_G=1/R_3$ in our numerical tests) that controls the gauge
condition, Eq.~(\ref{e:DampendHarmonicGauge}), used in our evolutions.
The modes responsible for this somewhat weaker gauge instability have
spatial structures determined by the various $k=1$ three-sphere
harmonics.  This instability can also be 
suppressed, therefore, by
constructing the appropriate $k=1$ mode-damping forces.

The $k=1$ parts of the Einstein-Klein-Gordon static solution have
$0=\psi^{1\ell m}(t) =f^{1\ell m}(t) =\psi_{tt}^{1\ell m}(t)
=f_{tt}^{1\ell m}(t) =\psi_{tj}^{1\ell m}(t) = f_{tj}^{1\ell m}(t) =
\phi^{1\ell m}(t) = f_\phi^{1\ell m}(t)$.  The evolution equations can
therefore be modified to drive the dynamical solution toward the state
having no $k=1$ three-sphere harmonic content by adding
the following mode-damping forces:
\begin{eqnarray}
&&
\!\!\!\!\!\!
\mathcal{D}f_{ab}^{\,1\ell m}\equiv-
 \frac{Y^{1\ell m}}{3R^3_3}\left[ f^{1\ell m}(t)  
+ \eta_G  \psi^{1\ell m}(t)\right]\tilde g_{ab}\nonumber\\
&&\quad-\frac{\hat t_a\tilde\nabla_bY^{1\ell m}
+\hat t_b\tilde\nabla_aY^{1\ell m}}{3R_3}
\left[ f^{1\ell m}_{tj}(t)+\eta_G  \psi^{1\ell m}_{tj}(t)\right],
\label{e:psi1lmDamping}
\nonumber\\
&&\quad-\frac{Y^{1\ell m}}{R^3_3}\left[ f_{tt}^{1\ell m}(t) 
+\eta_G \psi_{tt}^{1\ell m}(t)\right]\hat t_a \hat t_b,\\
&&
\!\!\!\!\!\!
\mathcal{D}f_\phi^{\,1\ell m}\equiv
- \frac{Y^{1\ell m}}{R_3^3}
\left[ f_\phi^{1\ell m}(t) +
\eta_S \phi^{k\ell m}(t)\right].
\label{e:phi1lmDamping}
\end{eqnarray}
Similar forces could be constructed to suppress the $k=1$ dynamics in
the evolution equations for $\Pi_{ab}$ and $\Pi_\phi$.  Such forces are
not needed to control the growth of this rather weak $k=1$
instability, however, so a minimalist approach has been followed by setting
$0=\mathcal{D}F_{ab}^{\,1\ell m} = \mathcal{D}F_\phi^{\,1\ell m}$.

Combining the $k=0$ damping forces from
Eqs.~(\ref{e:psi000Damping})--(\ref{e:Piphi000Damping}) with the $k=1$
forces from Eqs.~(\ref{e:psi1lmDamping}) and (\ref{e:phi1lmDamping})
gives the needed composite mode-damping forces:
\begin{eqnarray}
\mathcal{D}f_{ab}&=&\mathcal{D}f_{ab}^{\,000}+\sum_{\ell=0}^1
\sum_{m=-\ell}^\ell\mathcal{D}f_{ab}^{1\ell m},\label{e:psidotDamping}\\
\mathcal{D}F_{ab}&=&\mathcal{D}F_{ab}^{\,000},\label{e:PidotDamping}\\
\mathcal{D}f_\phi&=&\mathcal{D}f_\phi^{\,000}
+\sum_{\ell=0}^1\sum_{m=-\ell}^\ell\mathcal{D}f_\phi^{1\ell m},
\label{e:phidotDamping}\\
\mathcal{D}F_\phi&=&\mathcal{D}F_\phi^{\,000}.\label{e:PhiphidotDamping}
\end{eqnarray}
The resulting modified Einstein-Klein-Gordon evolution system
suppresses the dynamics in the $k=0$ three-sphere harmonic components
of $\psi_{ab}$, $\Pi_{ab}$, $\phi$ and $\Pi_\phi$ according to
Eqs.~(\ref{e:K0psiDamping})--(\ref{e:K0PiphiDamping}).  In addition,
the modified system also suppresses the dynamics in the
$k=1$ three-sphere harmonic components $\psi_{ab}$ and $\phi$
in the following way:
\begin{eqnarray}
\partial_t\psi_{tt}^{1\ell m}(t) &=& -\eta_G\psi_{tt}^{1\ell m}(t),\\
\partial_t\psi_{tj}^{1\ell m}(t) &=& -\eta_G\psi_{tj}^{1\ell m}(t),\\
\partial_t \psi^{1\ell m}(t) &=& -\eta_G\psi^{1\ell m}(t),\\
\partial_t\phi^{1\ell m}(t) &=& -\eta_S\phi^{1\ell m}(t).
\end{eqnarray}

\begin{figure}
\includegraphics[width=3.4in]{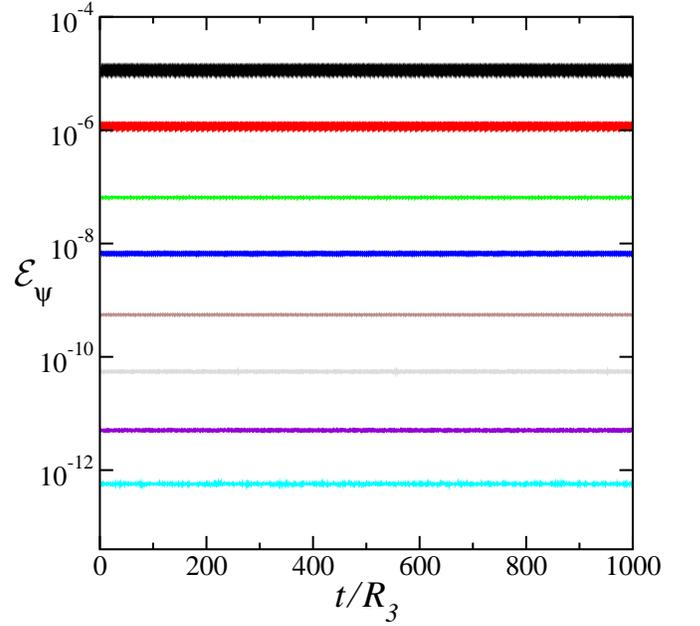}
\caption{\label{f:LpPsiErrJ1} Errors in the metric $\psi_{ab}$ for
  evolutions (including mode-damping forces) of initial data for the
  Einstein-Klein-Gordon static solution.  Numerical resolutions are
  the same as those shown in
  Figs.~\ref{f:EinsteinStaticNoDampErrPsi}--\ref{f:EinsteinStaticNoDampGhCe}.}
\end{figure}

\begin{figure}
\includegraphics[width=3.4in]{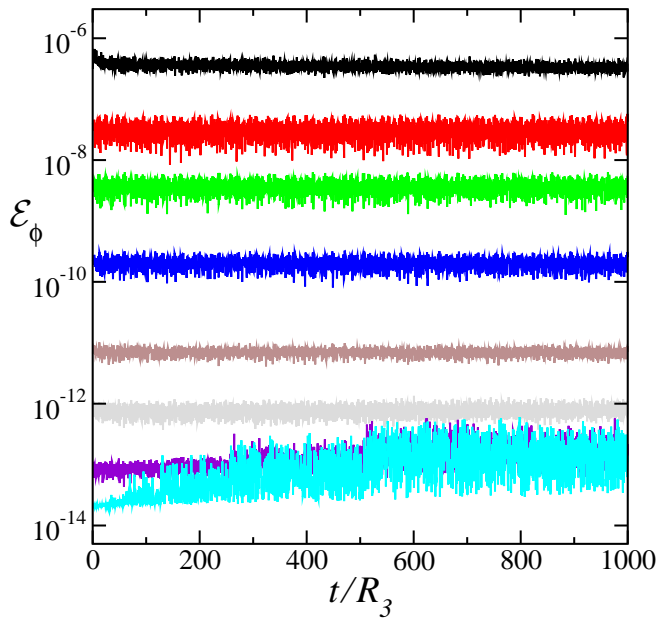}
\caption{\label{f:LpSwPsiErrJ1} Errors in the complex Klein-Gordon
  scalar field $\phi$ for evolutions (including mode-damping forces)
  of initial data for the Einstein-Klein-Gordon static solution.
  Numerical resolutions are the same as those shown in
  Figs.~\ref{f:EinsteinStaticNoDampErrPsi}--\ref{f:EinsteinStaticNoDampGhCe}.}
\end{figure}

\begin{figure}
\includegraphics[width=3.4in]{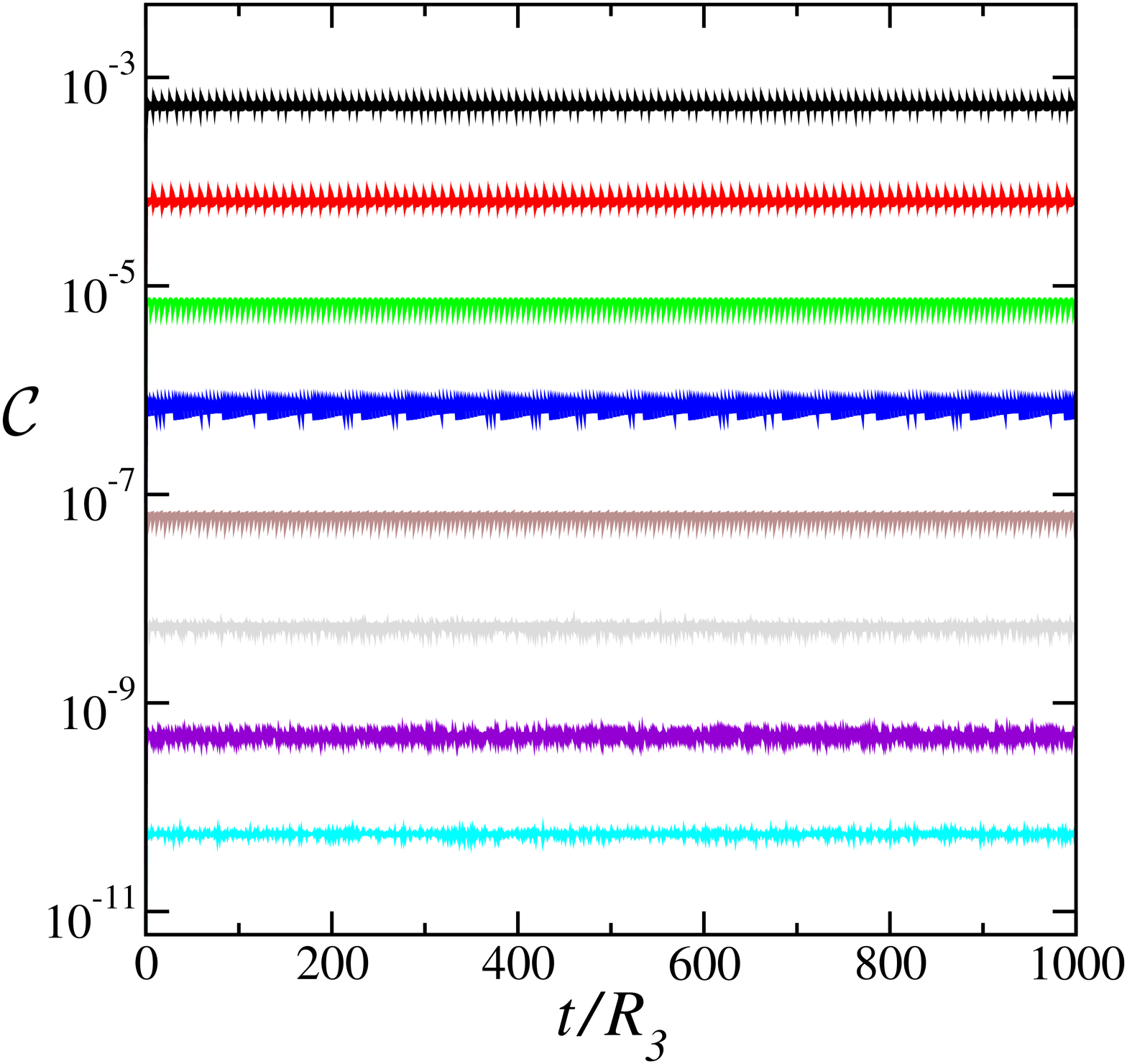}
\caption{\label{f:LpGhCeJ1} Constraint norm ${\cal C}$ for evolutions
  (including mode-damping forces) of initial data for the
  Einstein-Klein-Gordon static solution.  Numerical resolutions are
  the same as those shown in
  Figs.~\ref{f:EinsteinStaticNoDampErrPsi}--\ref{f:EinsteinStaticNoDampGhCe}.}
\end{figure}

The second numerical test of our implementation of the multicube
method evolves the coupled Einstein and Klein-Gordon evolution
equations, modified with the $k=0$ and $k=1$ mode-damping forces.  The
initial data used for these evolutions are those of the static
Einstein-Klein-Gordon universe solution, described in detail in
Sec.~\ref{s:EinsteinStatic}.  Figures~\ref{f:LpPsiErrJ1} and
\ref{f:LpSwPsiErrJ1} illustrate the errors in the metric $\psi_{ab}$
and the Klein-Gordon scalar field $\phi$, as measured by the
quantities $\mathcal{E}_\psi$ and $\mathcal{E}_\phi$ defined in
Eqs.~(\ref{e:EpsiDef}) and (\ref{e:EphiDef}), respectively.
Figure~\ref{f:LpGhCeJ1} illustrates the constraint norm $\mathcal{C}$
defined in Eq.~(\ref{e:CErrorDef}) for this test.  These results show
that the mode-damping forces are effective in suppressing the $k=0$ and
the $k=1$ instabilities that appeared in our earlier tests.  
The light-crossing time of the $S^3$ geometry 
is $2\pi R_3$, so these results
demonstrate numerical stability and convergence for about 160 
light-crossing times of the solution.

\begin{figure}
\includegraphics[width=3.4in]{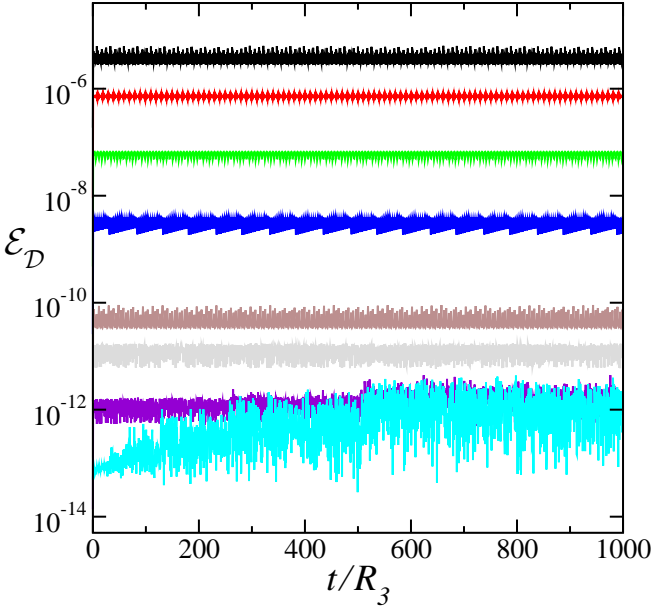}
\caption{\label{f:LpModeDampingJ1} Norm of the mode-damping forces,
  $\mathcal{E_D}$, for evolutions (including mode-damping forces)
  of initial data for the
  Einstein-Klein-Gordon static solution. Numerical resolutions are the
  same as those shown in
  Figs.~\ref{f:EinsteinStaticNoDampErrPsi}--\ref{f:EinsteinStaticNoDampGhCe}.}
\end{figure}

The results shown in
Figs.~\ref{f:LpPsiErrJ1}--\ref{f:LpGhCeJ1} demonstrate that the
constraints of the Einstein-Klein-Gordon evolution 
system are satisfied, and that the numerical
solution converges to the Einstein-Klein-Gordon static universe
solution.  These results do not demonstrate, however, that the
physical Einstein-Klein-Gordon equations are actually satisfied.  The
mode-damping forces, $\mathcal{D} f_{ab}$, $\mathcal{D} F_{ab}$,
$\mathcal{D} f_{\phi}$, and $\mathcal{D} F_{\phi}$ must be measured to
confirm that.  We measure the 
sizes of these mode-damping forces 
with the quantity $\mathcal{E_D}$, defined as the integral norm of
each component of each mode-damping force:
\begin{eqnarray}
\!\!\!\!\!\!\!\!\!\!
    \mathcal{E}_\mathcal{D}^{\,2} &\equiv& 
    \frac{\int m^{ab}m^{cd}\,\mathcal{D}f_{ac}\,
          \mathcal{D}f_{bd}\,\sqrt{g}\, d^{\,3}x}
         {\int \mu^2\, m^{ab}m^{cd}\,\psi_{ac}\psi_{bd}\, \sqrt{g}\, d^{\,3}x}
         \nonumber\\ 
         &&+\frac{\int m^{ab}m^{cd}\,\mathcal{D}F_{ac}\,
          \mathcal{D}F_{bd}\,\sqrt{g}\, d^{\,3}x}
         {\int \mu^4\, m^{ab}m^{cd}\,\psi_{ac}\psi_{bd}\, \sqrt{g}\, d^{\,3}x}
         \nonumber\\
         &&+\frac{\int |\mathcal{D}f_\phi|^2\, \sqrt{g}\, d^{\,3}x}
         {\int \mu^2 |\phi |^2\, \sqrt{g}\, d^{\,3}x}
         +\frac{\int |\mathcal{D}F_\phi|^2\, \sqrt{g}\, d^{\,3}x}
         {\int \mu^4 |\phi |^2\, \sqrt{g}\, d^{\,3}x}.
\end{eqnarray} 
The factors of $\mu$ (the fundamental scalar field oscillation
frequency) in this expression are used as 
characteristic time scales in the denominators
to make $\mathcal{E_D}$ dimensionless.
Figure~\ref{f:LpModeDampingJ1} shows that the mode-damping forces
converge to zero as the numerical resolution is
increased, so our numerical solution
also solves the 
unmodified {\sl physical} Einstein-Klein-Gordon evolution 
equations in this limit.
Consequently, the results shown in
Figs.~\ref{f:LpPsiErrJ1}--\ref{f:LpModeDampingJ1} 
demonstrate that
our implementation of the multicube method for solving Einstein's
equation on manifolds with nontrivial spatial  
topologies is stable and
numerically convergent even for very long-time-scale evolutions.

\section{Perturbed Einstein-Klein-Gordon Static Universe}
\label{s:PerturbedEinsteinStatic}

The numerical tests of the Einstein-Klein-Gordon evolution system
described in Sec.~\ref{s:ModeDamping} confirm that our implementation
of the multicube method for solving Einstein's equation 
described in 
Secs.~\ref{s:ReviewMultiCubeMethod} and \ref{s:CovariantEinsteinSystem} 
is basically correct and free of
numerical instabilities even on rather long time scales.  Those
numerical tests were limited, however, by the fact that the
Einstein-Klein-Gordon static universe solution is time independent and
its spatial structure is extremely simple.  In this section we address
these limitations by carrying out a third, more challenging, set of
numerical tests of the multicube methods by performing long-time-scale
evolutions of complicated time-dependent perturbations of the
Einstein-Klein-Gordon static universe solution.  We study these
perturbed solutions analytically in
Sec.~\ref{s:AnalyticalPerturbations} and numerically in
Sec.~\ref{s:NumericalTests}.  The 
results demonstrate that our
numerical nonlinear Einstein-Klein-Gordon code successfully evolves
complicated dynamical solutions having significant spatial structures.
We show that these numerical solutions converge to solutions of the
Einstein-Klein-Gordon evolution system that agree with the analytical
predictions.

\subsection{Analytical perturbations}
\label{s:AnalyticalPerturbations}

In this section we derive analytically the general solutions to the
coupled Einstein and Klein-Gordon equations for perturbations about
the Einstein-Klein-Gordon static universe solution.  Write the
spacetime metric $\psi_{ab}$ and the scalar field $\phi$ for this
perturbed solution as
\begin{eqnarray}
\psi_{ab}&=&\psi_{ab}^0+\delta\psi_{ab},\\
\phi &=& \phi_0\,e^{i\mu t}+\delta\phi,
\end{eqnarray}
where $\psi_{ab}^0$ and $\phi_0\,e^{i\mu t}$ are the ``background''
metric and scalar fields of the Einstein-Klein-Gordon static universe
solution.  The background metric $\psi^0_{ab}$ is identical to the
reference metric $\tilde\psi_{ab}$ used to fix the differential
structure in our multicube representation of $S^3$.
We will therefore refer to the background metric as $\tilde\psi_{ab}$.  
The evolution equations for the perturbations, $\delta\psi_{ab}$ and
$\delta\phi$, are obtained by linearizing the coupled
Einstein-Klein-Gordon equations about this background.  The perturbed
Ricci tensor is given by
\begin{eqnarray}
\delta R_{ab} &=& -\frac{1}{2} \tilde\nabla^c\tilde\nabla_c \delta\psi_{ab}
-\tilde\nabla_{(a}\delta H_{b)} + \frac{2}{R_3^2}\,\tilde g^c{}_{(a}\delta\psi_{b)c}
\nonumber\\
&&-\frac{1}{R_3^2}\left(\tilde g^{cd}\tilde g_{ab}
-\tilde g^c{}_{(a}\tilde g^d{}_{b)}\right)\delta\psi_{cd},
\label{e:PerturbedRicci}
\end{eqnarray}
where $\tilde \nabla_a$ is the covariant derivative associated with
the background metric $\tilde\psi_{ab}$, and $\tilde
g_{ab}=\tilde\psi_{ab}+ \tilde\nabla_a t\tilde \nabla_b t$ is the
background spatial metric.  The perturbed Einstein equation 
is given
by
\begin{eqnarray}
\delta R_{ab} = \left(\Lambda - 4\pi T_0\right)\delta\psi_{ab} 
+8\pi \left(\delta T_{ab} - \half\tilde\psi_{ab} \delta T\right),
\end{eqnarray}
where $\Lambda=1/R_3^2$ and $4\pi T_0=-1/R^2_3$ are the cosmological
constant and trace of the stress tensor from the background spacetime,
respectively,
and $\delta T_{ab}$ and $\delta T = \tilde\psi^{ab}\delta T_{ab}
-T_0^{ab}\delta\psi_{ab}$ are the perturbed stress-energy tensor and
its trace.  For the Einstein-Klein-Gordon system, the perturbed
stress-energy tensor is given by
\begin{eqnarray}
&&\!\!\!\!\!\!
\delta T_{ab}-\half\tilde\psi_{ab}\delta T =
\half\mu^2\left(\phi_0e^{i\mu t}\delta\phi^*
+\phi_0^*e^{-i\mu t}\delta\phi\right)\tilde\psi_{ab}
\nonumber\\
&& \quad
+i\mu\phi_0 e^{i\mu t}\tilde\nabla_{(a}\delta\phi^*\tilde\nabla_{b)}t
-i\mu\phi_0^* e^{-i\mu t}\tilde\nabla_{(a}\delta\phi\tilde\nabla_{b)}t.
\qquad
\end{eqnarray}
The perturbed Klein-Gordon equation for this system is given by
\begin{eqnarray}
0=\tilde\nabla^a\tilde\nabla_a \delta\phi - \mu^2\delta\phi
+\mu^2\phi_0 e^{i\mu t}\delta\psi_{tt}.
\end{eqnarray}
The perturbed damped harmonic gauge condition for this system is
given by
\begin{eqnarray}
0=\tilde\nabla^b\delta\psi_{ba}-\half\tilde\psi^{bc}\tilde\nabla_a
\delta\psi_{bc}-\mu_G\tilde g_a{}^b\delta\psi_{bt}.
\label{e:PerturbedGauge}
\end{eqnarray}
The perturbations of the Einstein-Klein-Gordon static solution are
determined by solving the linearized system,
Eqs.~(\ref{e:PerturbedRicci})--(\ref{e:PerturbedGauge}),
for $\delta\psi_{ab}$ and $\delta\phi$.

These perturbed Einstein-Klein-Gordon equations can be decoupled into
separate equations for the scalar, vector, and tensor degrees of
freedom of the system.  To accomplish this, the perturbed metric
$\delta\psi_{ab}$ is decomposed into 
two scalars (under spatial coordinate
transformations) $\delta\psi_{tt}$ and $\delta\psi =
\tilde\psi^{ij}\delta\psi_{ij}$, one vector $\delta\psi_{jt}$, and one
trace-free tensor $\delta\bar\psi_{ij}=
\delta\psi_{ij}-\frac{1}{3}\tilde\psi_{ij}\delta\psi$. 
These fields can then be represented as linear combinations
of the appropriate scalar, vector, and tensor harmonics on the
three-sphere (as described in Appendix~\ref{s:AppendixB}).  Since the
background Einstein-Klein-Gordon solution is static, the solutions to
the perturbation equations can be expressed as linear combinations of
modes, i.e., solutions having time dependence $e^{i\omega t}$.

We first discuss the modes corresponding to the scalar degrees of
freedom of the system.  The perturbations of $\delta\psi_{ab}$ and
$\delta \phi$ for a general scalar mode can be written in the form
\begin{eqnarray}
\delta\psi_{tt} &=& \Re\left[A_{tt} Y^{k\ell m}e^{i\omega_S t}\right],\\
\delta\psi_{tj} &=& \Im\left[A_{tj}Y^{k\ell m}_{(0)\,j} e^{i\omega_St}\right],\\
\delta\psi &=& \Re\left[A_{{\psi}}Y^{k\ell m}e^{i\omega_S t}\right],\\
\delta\bar\psi_{{jk}} &=& \Im\left[A_{\bar{jk}}Y^{k\ell m}_{(3)\,jk}e^{i\omega_S t}
\right],\\
\delta\phi &=& \phi_0e^{i\mu t} \left[A_\phi^+ Y^{k\ell m} e^{i\omega_St}
+A_\phi^{-*} Y^{k\ell m*} e^{-i\omega_St}\right],\qquad
\end{eqnarray}
where $A_{tt}$, $A_{tj}$, $A_{\psi}$, $A_{\bar{jk}}$, $A^+_{\phi}$,
and $A^-_{\phi}$ are complex constants; $Y^{k\ell m}$,
$Y_{(0)\,j}^{k\ell m}$, and $Y^{k\ell m}_{(3)\,jk}$ are the scalar,
vector, and tensor harmonics on $S^3$ defined in
Appendix~\ref{s:AppendixB}; $\omega_S$ is the frequency of the mode;
and $\Re (Z)$ and $\Im (Z)$ denote the real and imaginary parts of a
quantity $Z$, respectively.  The perturbed Einstein-Klein-Gordon
equations for these perturbations become a system of linear algebraic
equations
for the amplitudes $A_{tt}$, ....  These linear equations have
solutions whenever the frequency $\omega_S$ is one of the mode
eigenfrequencies of the system.  For these values of $\omega_S$
the general solution to the perturbation equations can be written as
\begin{eqnarray}
A_{tt} &=& A_S^{k\ell m},
\label{e:scalarmodeeqtt}\\
A_{\psi} &=& -A_S^{k\ell m}
-\frac{16k(k+2)\mu^2R_3^2}{Q}A_S^{k\ell m},\\
A_\phi^+  &=& -\frac{\mu^2R_3^2}{2[\omega_S(\omega_S+2\mu)R_3^2-k(k+2)]}
A_S^{k\ell m},\\
A_\phi^-  &=& -\frac{\mu^2R_3^2}{2[\omega_S(\omega_S-2\mu)R_3^2-k(k+2)]}
A_S^{k\ell m},
\label{e:scalarmodephi-}
\\
A_{tj} &=& -\frac{8\mu^2\omega_SR_3^4}{Q}A_S^{k\ell m},
\label{e:scalarmodepsitj}\\
A_{\bar{jk}} &=& -\frac{16\mu_G\mu^2\omega_SR_3^6}{Q[\omega_S^2R_3^2+4-k(k+2)]}
A_S^{k\ell m},\label{e:scalarmodebarjk}
\end{eqnarray}
where $A_S^{k\ell m}$ is the complex constant that sets the amplitude
of the scalar mode, and $Q$ is defined by
\begin{eqnarray}
Q&=& [\omega_S(\omega_S-i\mu_G)R_3^2+4-k(k+2)]\nonumber\\
&&\qquad\times\left\{[\omega_S^2R_3^2-k(k+2)]^2-4\mu^2\omega_S^2R_3^4\right\}.
\qquad
\end{eqnarray}  
The allowed eigenfrequencies of these modes break up into
three distinct families, defined by 
\begin{eqnarray}
&&(\omega_S^0R_3)^2 = k(k+2),\label{e:scalarmodefreqs00}\\
&&(\omega_S^\pm R_3)^2 = k(k+2)+2(\mu^2R_3^2  -1)\nonumber \\
&&\qquad\quad
 \pm 2\sqrt{(\mu^2R_3^2-1)^2 + \left[k(k+2)+1\right]\mu^2 R_3^2}.\qquad
\label{e:scalarmodefreqspm}
\end{eqnarray}
It is straightforward to show that $(\omega_S^\pm R_3)^2> 0$ when
$k\geq 2$ and $8\geq \mu^2R_3^2$, so the generic scalar modes are
stable in these cases.

The scalar modes for the cases $k=0$ and $k=1$ are somewhat
exceptional and must be calculated separately.  For the $k=0$ case,
the vector and tensor harmonics, $Y^{k\ell m}_{(0)\,j}$ and $Y^{k\ell
  m}_{(3)\,ij}$, both vanish, so the mode amplitudes $A_{tj}$ and
$A_{\bar{ij}}$ are effectively zero.  The mode amplitudes of the
remaining scalar degrees of freedom, $A_{tt}$, $A_{\psi}$, $A_\phi^+$,
and $A_\phi^-$, are given by the expressions in
Eqs.~(\ref{e:scalarmodeeqtt})--(\ref{e:scalarmodephi-}) with $k=0$,
but there are only two independent mode frequencies in this case:
\begin{eqnarray}
(\omega_S^\pm R_3)^2 &=& 2\mu^2R_3^2 -2\pm 2\sqrt{\mu^4R_3^4 - \mu^2
    R_3^2+1}.\qquad
\label{e:scalarmodefreqs0}
\end{eqnarray}
One of these has an imaginary frequency, $(\omega_S^-R_3)^2<0$, and
therefore represents an unstable mode of the Einstein-Klein-Gordon
system.  The instability seen in the numerical evolution discussed in
Sec.~\ref{s:EinsteinStatic} has a growth rate that matches with great
accuracy the analytical rate predicted by this unstable $k=0$ mode
frequency, $\omega^-_S$.  There is also a degenerate exceptional $k=0$
mode having $\omega_SR_3=0$.  This mode has $A_{tt}=A_{\psi}=0$ and
$A_\phi^+=-A_\phi^-$.  This exceptional mode does not excite the
gravitational field at all and appears to be a kind of gauge mode
associated with the phase of the complex scalar field $\phi$.

The other exceptional scalar modes are those with $k=1$.  In this case
the tensor harmonics $Y^{k\ell m}_{(3)\,ij}$ vanish identically, so in
effect $A_{\bar{jk}}=0$. Repeating the mode calculation
gives the expressions in
Eqs.~(\ref{e:scalarmodeeqtt})--(\ref{e:scalarmodepsitj}) with $k=1$.
There are, however, a smaller number of mode frequencies in this case:
$$(\omega_S^\pm R_3)^2 = 3 + 2\mu^2R_3^2 \pm 2\mu^2 R_3^2,$$ both of which
satisfy $(\omega_S^\pm R_3)^2>0$ and are therefore stable.  In addition,
there are two other $k=1$ modes that have somewhat different mode
structures.  For these modes,
\begin{eqnarray}
A_{tt}&=&A^+_{\phi}=A^-_{\phi}=0,\\
A_{\psi} &=& 6(\omega_S-i\mu_G)R_3A^{k\ell m}_S,\\
A_{tj}&=&A_S^{k\ell m}.
\end{eqnarray}
The frequencies of these exceptional $k=1$ modes are given by
\begin{eqnarray}
\omega_S^\pm R_3 = \frac{i}{2}\left(\mu_GR_3\pm \sqrt{4+\mu_G^2R_3^2}\right).
\end{eqnarray}
One of these modes is a nonoscillatory damped mode, while the other
mode is unstable.  The instability seen in the preliminary numerical
evolution discussed in Sec.~\ref{s:ModeDamping} has a growth rate that
matches the analytical rate predicted by this ($k=1$)-mode frequency
$\omega^-_S$.  This exceptional $k=1$ mode does not excite the
Klein-Gordon scalar field at all and appears to be associated with
the coordinate gauge freedom of the gravitational field.

The Einstein-Klein-Gordon perturbation equations also admit mode
solutions that represent the vector degrees of freedom of the
gravitational field.  The modes representing these vector degrees of
freedom can be written quite generally as
\begin{eqnarray}
\!\!\!\!\!
\!\!\!\!\!
\delta\psi_{tj}\!\! &=&\!\! \Re\left\{i\omega_V\left[A^{k\ell m}_{V(1)} Y^{k\ell m}_{(1)\,j}
  +A^{k\ell m}_{V(2)}Y^{k\ell m}_{(2)\,j}\right] e^{i\omega_Vt}\right\},\\
\!\!\!\!\!
\!\!\!\!\!
\delta\bar\psi_{jk}\!\! &=&\!\! \Re\left\{2\left[A^{k\ell m}_{V(1)} Y^{k\ell m}_{(1)\,jk}
  +A^{k\ell m}_{V(2)}Y^{k\ell m}_{(2)\,jk}\right] e^{i\omega_Vt}\right\}.
\end{eqnarray}
Here, $A^{k\ell m}_{V(1)}$ and $A^{k\ell m}_{V(2)}$ are (complex)
constants; and $Y^{k\ell m}_{(1)\,j}$, $Y^{k\ell m}_{(2)\,j}$,
$Y^{k\ell m}_{(1)\,jk}$, and $Y^{k\ell m}_{(2)\,jk}$ are the type-1 and type-2
vector and tensor harmonics defined in Eqs.~(\ref{e:Yklm(1)i}),
(\ref{e:Yklm(2)i}), (\ref{e:Yklm(1)ij}), and (\ref{e:Yklm(2)ij}) in
Appendix~\ref{s:AppendixB}.  These harmonics are defined only for
$k\geq 1$.  The perturbed Einstein-Klein-Gordon equations admit
solutions of this type for arbitrary values of the mode amplitudes,
$A^{k\ell m}_{V(1)}$ and $A^{k\ell m}_{V(2)}$, whenever the frequency
$\omega_V$ satisfies the vector-mode eigenfrequency condition 
\begin{eqnarray}
(\omega_V-i\mu_G/2)^2R_3^2=k(k+2)-3-\mu_G^2R_3^2/4.
\end{eqnarray}
The quantity $\mu_G>0$ that appears in these expressions is the
harmonic gauge damping factor defined in
Eq.~(\ref{e:DampendHarmonicGauge}).  
The frequencies of these modes
are complex with non-negative imaginary parts, so these vector modes
are all stable.  These vector modes appear to 
be associated with 
the spatial 
coordinate gauge degrees of freedom of the system.

Finally, there is a set of modes that represent the tensor degrees of
freedom of the system.  The two tensor degrees of freedom are the
trace-free, $\delta\bar\psi_{jk} = \delta\psi_{jk}-
\frac{1}{2}\psi_{0\,jk}\psi_0^{rs}\delta\psi_{rs}$, and
transverse, $\nabla^k\delta\bar\psi_{jk}=0$, parts of the metric
perturbation.  The general form for these tensor modes is given by
\begin{eqnarray}
\!\!\!\!\!\!
\delta\bar\psi_{jk} &=& \Re\left\{\left[A^{k\ell m}_{T(4)} Y^{k\ell m}_{(4)\,jk}
  +A^{k\ell m}_{T(5)}Y^{k\ell m}_{(5)\,jk}\right] e^{i\omega_Tt}\right\},
\end{eqnarray}
where $A^{k\ell m}_{T(5)}$ and $A^{k\ell m}_{T(5)}$ are constants, and
$Y^{k\ell m}_{(4)\,jk}$ and $Y^{k\ell m}_{(5)\,jk}$ are the type-4 and type-5
tensor harmonics defined in Eqs.~(\ref{e:Yklm(4)ij})
and~(\ref{e:Yklm(5)ij}) in Appendix~\ref{s:AppendixB}.  These tensor
harmonics exist only for $k\geq 2$ and $\ell\geq 2$.  The perturbed
Einstein-Klein-Gordon equations for these modes are satisfied for
arbitrary (small) values of the complex constants $A^{k\ell m}_{T(5)}$
and $A^{k\ell m}_{T(5)}$, as long as the frequency $\omega_T$ satisfies
the tensor-mode eigenfrequency condition
\begin{eqnarray}
\omega_T^2R_3^2=k(k+2).
\end{eqnarray}
These frequencies are real, $\omega_T^2R_3^2>0$, so the 
transverse-traceless tensor modes are all stable.  
These tensor modes correspond
to the gravitational radiation degrees of freedom of the system.

We note that the modes of the Einstein-Klein-Gordon static universe
found in these analytical solutions are all stable, except for two
unstable modes.  These unstable $k=0$ and $k=1$ modes correspond
exactly to the unstable modes found in the numerical tests described
in Secs.~\ref{s:EinsteinStatic} and \ref{s:ModeDamping}.  This fact
provides additional (indirect) evidence that our numerical
implementation of the multicube method has been done correctly.

\subsection{Numerical tests}
\label{s:NumericalTests}

\begin{figure*}[tp!] 
  \centering
  \includegraphics[width=0.92\linewidth]{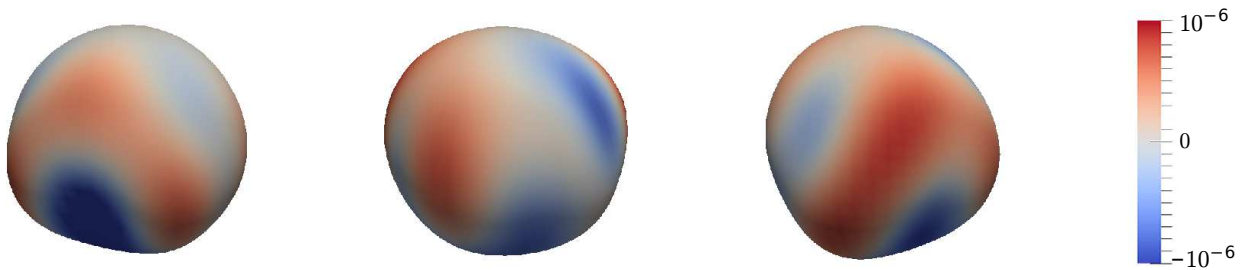}
\caption{\label{f:PerturbedLapse} Images of the $\delta\psi_{tt}$
  component of the metric perturbation, evaluated on the equatorial
  two-sphere, $\chi=\pi/2$ , of the perturbed Einstein-Klein-Gordon
  static solution.  These images represent the times $t=0$, $t=6 R_3$, and
  $t=12R_3$.  The color coding and distortion of the sphere represent
  the (scaled) magnitude of $\delta\psi_{tt}$.}
\end{figure*}

The third numerical test of our implementation of the multicube
method evolves initial data constructed from the analytical
perturbation solutions of the coupled Einstein-Klein-Gordon
evolution equations described in Sec.~\ref{s:AnalyticalPerturbations}.
We define the analytical metric, $\psi^\mathcal{A}_{ab}$, and 
scalar field, $\phi^\mathcal{A}$, solutions to be
\begin{eqnarray}
\psi^\mathcal{A}_{ab}&=&\tilde\psi_{ab}+\delta\psi_{ab},
\label{e:AnalyticPerturedMetric}\\
\phi^\mathcal{A}&=& \phi_0e^{i\mu t}+\delta\phi.
\label{e:AnalyticPerturebedScalarFields}
\end{eqnarray}
We construct the $\delta\psi_{ab}$ and $\delta\phi$ that appear in
these definitions by taking linear combinations of the scalar mode
solutions described in
Eqs.~(\ref{e:scalarmodeeqtt})--(\ref{e:scalarmodebarjk}).  We include
fifteen distinct scalar modes with spherical harmonic 
indices ranging from $k=2$ to $k=6$ and with a variety of 
values of the spherical harmonic indices $\ell$ and $m$.
The amplitudes $A^{k\ell m}_S$ of the individual modes used to
construct this solution are given in Table~\ref{t:TableI}.  Also
included in Table~\ref{t:TableI} is the choice of eigenfrequency
class for each mode, as defined in 
Eqs.~(\ref{e:scalarmodefreqs00}) and
(\ref{e:scalarmodefreqspm}).
\begin{table}[!b]
\caption{Amplitudes and frequency classes
  of the individual modes of the perturbed Einstein-Klein-Gordon
  system used to form the analytic perturbation solution 
  for the
  long-term stability tests shown in
  Figs.~\ref{f:LpPsiErrJ4}--\ref{f:LpModeDampingJ4}.
\label{t:TableI}}
\begin{tabular}{rrrccrl|rrrccrl}
\hline\hline
\,$k$ &\,\, $\ell$ &\,\, $m$ &\,\,\,& $A_S^{k\ell m}$
&\,\,\,& $\omega_S$\,\,\,
& \,\,$k$ &\,\, $\ell$ &\,\, $m$ &\,\,\,& $A_S^{k\ell m}$ 
&\,\,\,& $\omega_S$\,\\
\hline
2  &  2  &  2  &&  $1.0\times 10^{-6}$  && $\omega_S^0$
& 5 & 5 & 5 &&$4.0\times 10^{-7}$ && $\omega_S^0$\\
2 & 2 & -1 && $1.0\times 10^{-6}$ && $\omega_S^+$
& 5 & 5 & 4 &&$4.0\times 10^{-7}$ && $\omega_S^+$\\
2 & 1 & 1 && $1.0\times 10^{-6}$ && $\omega_S^-$
& 5 & 4 & -3 &&$4.0\times 10^{-7}$ && $\omega_S^-$\\
3 & 3 & -2 && $6.7\times 10^{-7}$ && $\omega_S^0$
& 6 & 6 & 6 &&$3.3\times 10^{-7}$ && $\omega_S^0$\\
3 & 3 & 1 && $6.7\times 10^{-7}$ && $\omega_S^+$
& 6 & 6 & -5 &&$3.3\times 10^{-7}$ && $\omega_S^+$\\
3 & 2 & 0 && $6.7\times 10^{-7}$ && $\omega_S^-$
& 6 & 5& 3 &&$3.3\times 10^{-7}$ && $\omega_S^-$\\
4 & 4 & -4 && $5.0\times 10^{-7}$ && $\omega_S^0$\\
4 & 4 & 3 && $5.0\times 10^{-7}$ && $\omega_S^+$\\
4 & 3 & -2 && $5.0\times 10^{-7}$ && $\omega_S^-$\\
\hline
\end{tabular}
\end{table}
The amplitudes of these modes were chosen to be about $10^{-6}$ (or
smaller) to ensure that the second-order (in amplitude) terms would be
comparable to the double-precision round-off errors in our numerical
evolutions.  We chose this particular mix of harmonics to produce a
solution having a complicated and interesting-looking dynamical
evolution.  Figure~\ref{f:PerturbedLapse} illustrates the metric
perturbation $\delta\psi_{tt}$ for this solution evaluated on the
equatorial two-sphere, $\chi=\pi/2$, of the three-sphere geometry.
The individual frames in Fig.~\ref{f:PerturbedLapse} illustrate this
field at times $t=0$, $t=6R_3$, and $t=12R_3$.  These times
(approximately one light-crossing time apart) do not correspond to any
natural period of the system, and are intended to illustrate the
complex, chaotic-looking dynamics produced by the chosen initial data.

\begin{figure}
\includegraphics[width=3.4in]{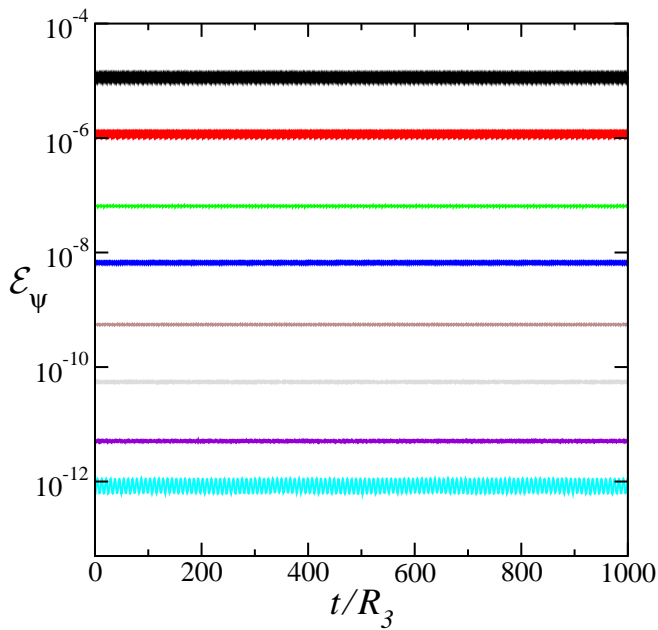}
\caption{\label{f:LpPsiErrJ4} Errors in the metric $\psi_{ab}$ for
  evolutions (including mode-damping forces) of initial data for the
  perturbed Einstein-Klein-Gordon solution.  Numerical resolutions are
  the same as those shown in
  Figs.~\ref{f:EinsteinStaticNoDampErrPsi}--\ref{f:EinsteinStaticNoDampGhCe}.}
\end{figure}

\begin{figure}
\includegraphics[width=3.4in]{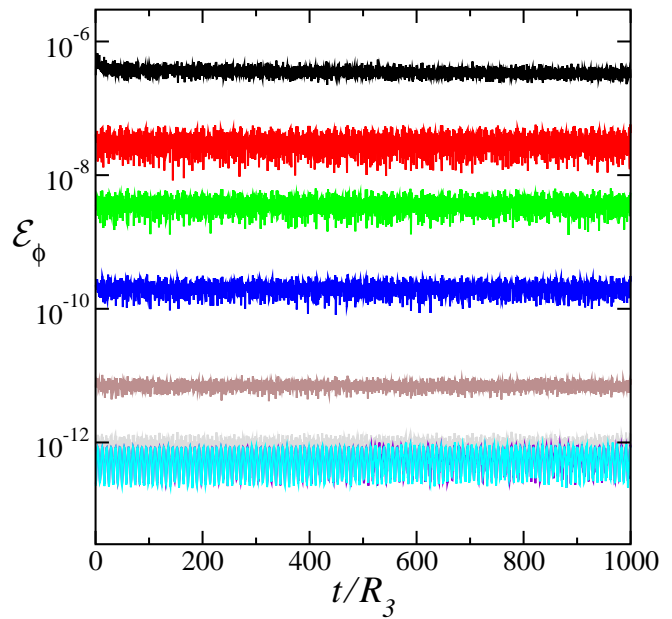}
\caption{\label{f:LpSwPsiErrJ4} Errors in the complex Klein-Gordon
  field $\phi$ for evolutions (including mode-damping forces) of
  initial data for the perturbed Einstein-Klein-Gordon solution.
  Numerical resolutions are the same as those shown in
  Figs.~\ref{f:EinsteinStaticNoDampErrPsi}--\ref{f:EinsteinStaticNoDampGhCe}.}
\end{figure}

\begin{figure}
\includegraphics[width=3.4in]{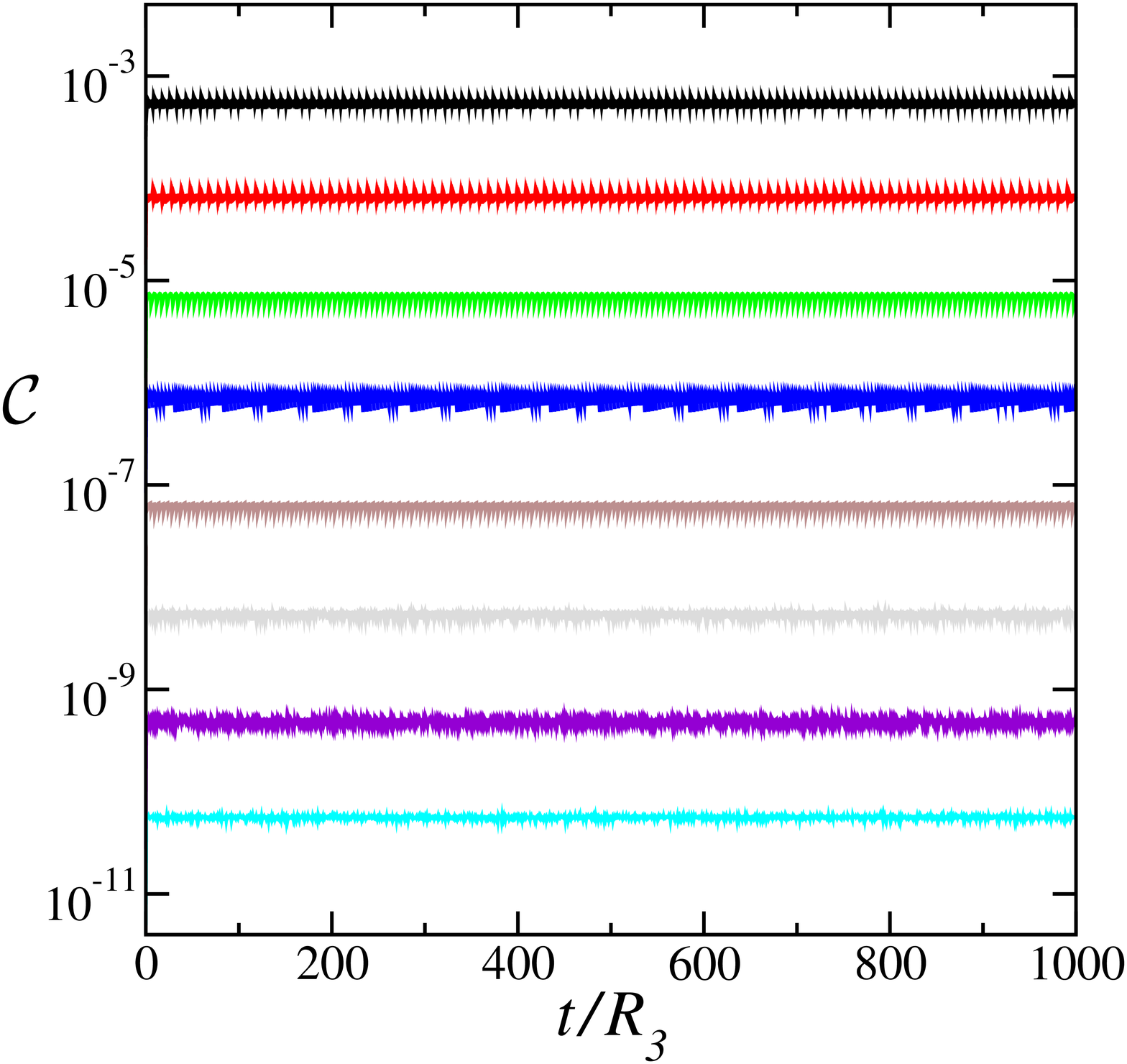}
\caption{\label{f:LpGhCeJ4} Constraint norm ${\cal C}$
for evolutions (including mode-damping forces) of
  initial data for the perturbed Einstein-Klein-Gordon solution.
  Numerical resolutions are the same as those shown in
  Figs.~\ref{f:EinsteinStaticNoDampErrPsi}--\ref{f:EinsteinStaticNoDampGhCe}.}
\end{figure}

\begin{figure}
\includegraphics[width=3.4in]{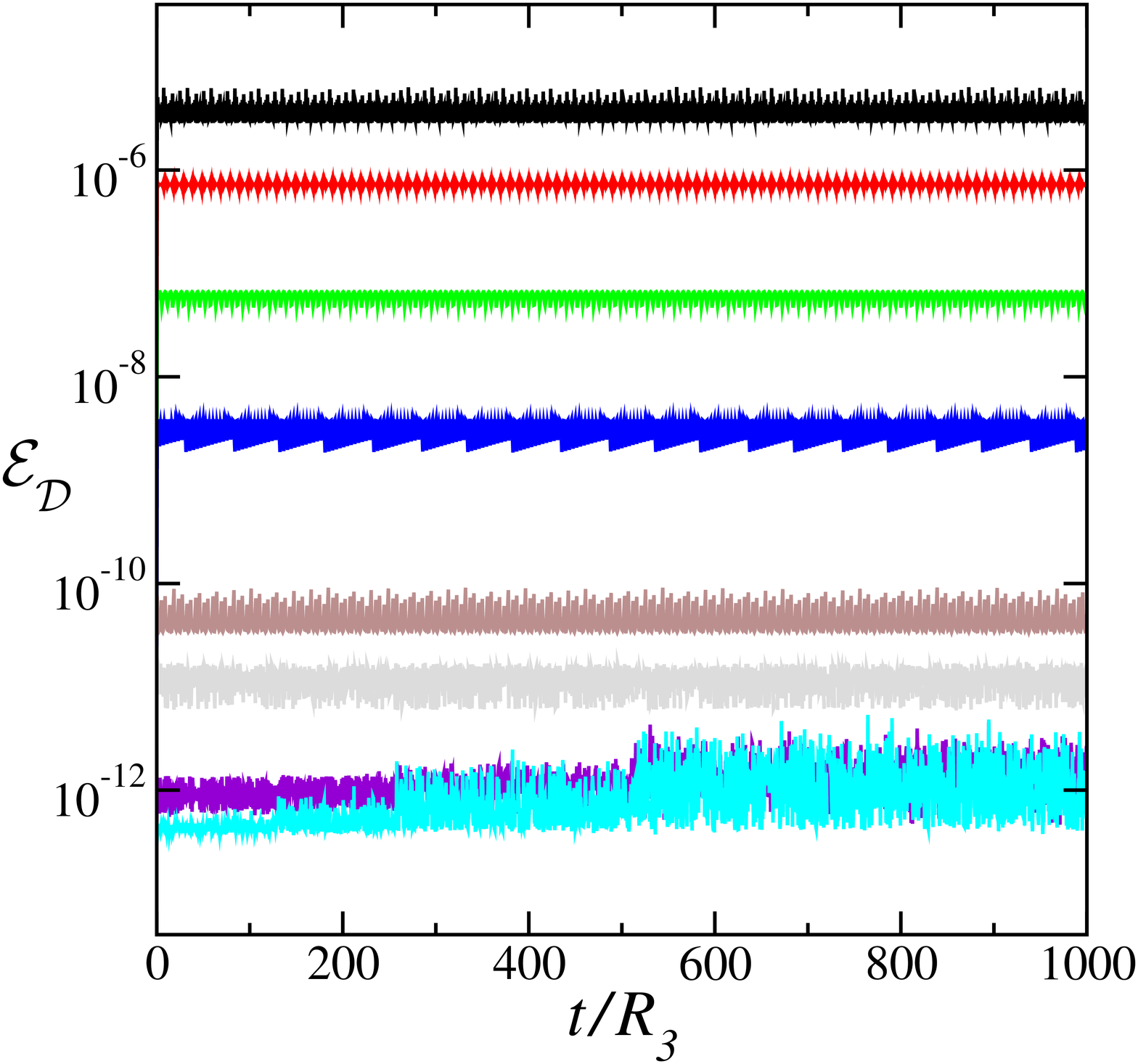}
\caption{\label{f:LpModeDampingJ4} Norm of the mode-damping forces,
  ${\cal E_D}$, for evolutions (including mode-damping forces) of initial
  data for the perturbed Einstein-Klein-Gordon solution.  Numerical
  resolutions are the same as those shown in
  Figs.~\ref{f:EinsteinStaticNoDampErrPsi}--\ref{f:EinsteinStaticNoDampGhCe}.}
\end{figure}

We use the analytical fields $\psi^\mathcal{A}_{ab}$ and
$\phi^\mathcal{A}$ defined in Eqs.~(\ref{e:AnalyticPerturedMetric})
and (\ref{e:AnalyticPerturebedScalarFields}) to construct initial data
for the Einstein-Klein-Gordon evolution system.  We evolve these data
numerically using the Einstein-Klein-Gordon equations that include the
unphysical mode-damping forces defined in
Eqs.~(\ref{e:psidotDamping})--(\ref{e:PhiphidotDamping}).
Figures~\ref{f:LpPsiErrJ4} and \ref{f:LpSwPsiErrJ4} illustrate the
differences between the numerically determined fields,
$\psi^\mathcal{N}_{ab}$ and $\phi^\mathcal{N}$, and the analytical
fields defined in Eqs.~(\ref{e:AnalyticPerturedMetric}) and
(\ref{e:AnalyticPerturebedScalarFields}).  These results show that the
numerical solutions converge toward the analytical solutions
until the size of
their differences approaches $10^{-12}$.  The analytical fields were
constructed from solutions to the first-order perturbation 
equations, and
so they are expected to contain errors at this level of accuracy.
Figures~\ref{f:LpGhCeJ4} and \ref{f:LpModeDampingJ4} show that the
constraints of the Einstein-Klein-Gordon system as well as the
unphysical mode-damping forces are 
numerically convergent (toward zero) in these  evolutions.  
These tests provide strong additional
evidence that our implementation of the multicube method for solving
Einstein's equation described in 
Secs.~\ref{s:ReviewMultiCubeMethod} and \ref{s:CovariantEinsteinSystem} is
correct and free from numerical instabilities. 

\section{Summary}
\label{s:Summary}

In this paper we extend 
the multicube method for solving
partial differential equations on manifolds with arbitrary spatial
topologies, developed in Ref.~\cite{LindblomSzilagyi2011a}, to allow us
to solve Einstein's equation on such manifolds.  We accomplish this by
developing in Sec.~\ref{s:CovariantEinsteinSystem} a new spatially
covariant first-order symmetric hyperbolic representation of
Einstein's equation.  This new representation is equivalent to the
standard noncovariant first-order generalized harmonic
representations (e.g., Ref.~\cite{Lindblom2006}) on manifolds with
spatial slices that can be embedded in $R^3$.  We test our
implementation of these multicube methods in the SpEC code (developed
by the SXS Collaboration, originally at Caltech and Cornell)
in Sec.~\ref{s:EinsteinStatic} by evolving initial data for a new
representation of the Einstein static universe metric on $R\times
S^3$.  Our representation uses a complex Klein-Gordon scalar field to
provide the energy density for this spacetime.  These numerical tests
reproduce with great precision
the well-known Eddington~\cite{Eddington1030} instability of
the Einstein static universe.  

We have tested the accuracy and the long-time-scale numerical stability
of our implementation of these multicube methods by adding unphysical
damping forces to Einstein's equation in Sec.~\ref{s:ModeDamping}.
These damping forces are designed to suppress the modes responsible
for the Eddington instability and to leave all the other dynamical
degrees of freedom of the system unchanged.  These long-time-scale
tests confirm stability and numerical convergence for about 160
light-crossing times of the $S^3$ geometry.  Finally, we have derived
analytical expressions for all of the modes of the
Einstein-Klein-Gordon static universe in
Sec.~\ref{s:PerturbedEinsteinStatic}.  We use these analytical
expressions to construct initial data for a complicated,
time-dependent spacetime having considerable spatial structure.  Our
numerical evolutions of these initial data converge toward the 
(small-amplitude) analytical perturbation solution, while the constraints and
mode-damping forces converge toward zero, as the spatial resolution is
increased. 

The numerical tests presented in this paper are all performed on the
manifold $R\times S^3$.  Nevertheless, we believe that these tests
confirm that the multicube methods described in
Secs.~\ref{s:ReviewMultiCubeMethod} and
\ref{s:CovariantEinsteinSystem} for solving Einstein's equation on
manifolds with arbitrary spatial topologies have been implemented
correctly.  In the multicube method, the equations are solved locally
within each cubic region $\cal{B}_A$ with boundary conditions,
cf. Sec.~\ref{s:BCHyperbolicSystems}, that guarantee that the solution
within each region corresponds to the desired global solution.  These
boundary conditions depend on the topology of the manifold only
through their dependence on the reference metric $\tilde\psi_{ab}$ and
the interface boundary maps $\Psi^{A\alpha}_{B\beta}$.  So while the
simulations presented here do not test reference metrics or interface
boundary maps for a wide range of manifolds with ``arbitrary''
topologies, they do verify that the basic structure of the boundary
conditions that would apply for arbitrary topologies has been
done correctly.


\acknowledgments

We thank Michael Holst, James Isenberg, Oliver Rinne, 
and Manuel Tiglio for helpful
discussions concerning this work.  We thank the KITP at the University
of California at Santa Barbara and the Mathematical Sciences
Center at Tsinghua University in Beijing, China, for their hospitality 
during the time
that a portion of this work was performed.  We also thank the Center
for Computational Mathematics at the University of California at San
Diego for providing access to their computer cluster on which the
numerical tests reported in this paper were performed.  This research
has been supported by a grant from the Sherman Fairchild Foundation
and by NSF Grants No.~PHY1005655 and No.~DMS1065438.


\appendix

\section{Covariant Einstein Constraints}
\label{s:AppendixA}

This appendix presents explicit expressions for the covariant
constraints of the Einstein evolution system derived in
Sec.~\ref{s:CovariantEinsteinSystem} in terms of the covariant
first-order dynamical fields $\psi_{ab}$, $\Pi_{ab}$, and $\Phi_{iab}$
and their spatial derivatives.  The primary constraint $\mathcal{C}_a$
of this system, defined in Eq.~(\ref{e:ConstraintGauge}), has the
following expression in terms of the first-order fields:
\begin{eqnarray}
{\cal C}_a &=& H_a + g^{ij}\Phi_{ija}
		 +t^b \Pi_{ba}
                  -\half g_a^i \psi^{bc}\Phi_{ibc}
		 -\half t_a \psi^{bc}\Pi_{bc}.\nonumber\\
\label{e:OneIndexConstraint}
\end{eqnarray}
The three-index constraint $\mathcal{C}_{iab}$, defined in
Eq.~(\ref{e:Constraint3index}), has the following expression:
\begin{eqnarray}
{\cal C}_{iab}=\tilde\nabla_i\psi_{ab}-\Phi_{iab}.
\label{e:ThreeIndexConstraint}
\end{eqnarray}
The spatially covariant analog of the Hamiltonian and momentum
constraints of more standard 3+1 representations of Einstein's
equation, $\mathcal{F}_a$, defined in Eq.~(\ref{e:FConstraintDef}),
has the following explicit representation in terms of
the first-order fields:
\begin{eqnarray}
\label{e:TimeDerivOfOneIndexConstraint}
{\cal F}_a &\equiv& 
\half g_a^i \psi^{bc}\tilde\nabla_i \Pi_{bc}
- g^{ij} \tilde\nabla_i \Pi_{ja}
- g^{ij} t^b \tilde\nabla_i \Phi_{jba}
\nonumber \\ &&
+ \half t_a \psi^{bc} g^{ij} \tilde\nabla_i \Phi_{jbc}
+ t_a g^{ij} \tilde\nabla_i H_j 
- g_a^i t^b \tilde\nabla_i H_b
\nonumber \\ &&
+ g_a^i \Phi_{ijb} g^{jk}\Phi_{kcd} \psi^{bd} t^c
- \half g_a^i \Phi_{ijb} g^{jk}
  \Phi_{kcd} \psi^{cd} t^b
\nonumber \\ &&
+ g^{ij} \Phi_{icd} \Phi_{jba} \psi^{bc} t^d
- \half t_a g^{ij} g^{mn} \Phi_{imc} \Phi_{njd}\psi^{cd}
\nonumber \\ &&
- \fourth  t_a g^{ij}\Phi_{icd}\Phi_{jbe}
   \psi^{cb}\psi^{de}
+ \fourth}{  t_a \Pi_{cd} \Pi_{be} 
   \psi^{cb}\psi^{de}
\nonumber \\ &&
+ \half t_a \Pi_{cd} \Pi_{be}\psi^{ce}
  t^d t^b
+ g_a^i \Phi_{icd} \Pi_{be} t^c t^b \psi^{de}
\nonumber \\ &&
- t^b g^{ij} \Pi_{b i} \Pi_{ja}
- \fourth  g_a^i \Phi_{icd} t^c t^d \Pi_{be}
  \psi^{be}
+ 2\Lambda t_a
\nonumber \\ &&
- g^{ij}\Phi_{iba} t^b \Pi_{je} t^e
- \half g^{ij}\Phi_{icd} t^c t^d \Pi_{ja}
-16\pi T_{ab}t^b 
\nonumber \\ &&
+\gamma_2\bigl(g^{id}{\cal C}_{ida}
-\half  g_a^i\psi^{cd}{\cal C}_{icd}\bigr)
-\Delta^b{}_{ac}t^c{\cal C}_b
\nonumber \\ &&
+2g^{ij}t^c\psi{}_{k(j}\tilde R^k{}_{a)ic}
-2\psi^{ij}t^b\tilde R^k{}_{ij(a}\psi{}_{b)k}
\nonumber \\ &&
-g_a{}^i\psi^{bd}t^c\psi_{j(b}\tilde R^j{}_{d)ic}
+t_a\psi^{bd}\psi^{ij}\tilde R^k{}_{ij(b}\psi{}_{d)k}.
\end{eqnarray}
Similarly, the two-index constraint, $\mathcal{C}_{ia}$, defined
in Eq.~(\ref{e:TwoIndexConstraintDef}), is given by
the expression
\begin{eqnarray}
\label{e:TwoIndexConstraint}
{\cal C}_{ia} &\equiv& g^{jk}\tilde\nabla_i \Phi_{jka} 
- \half g_a^j\psi^{cd}\tilde\nabla_i \Phi_{jcd} 
+ t^b \tilde\nabla_i \Pi_{ba}
\nonumber\\&&
- \half t_a \psi^{cd}\tilde\nabla_i\Pi_{cd}
+ \tilde\nabla_i H_a 
+ \half g_a^j \Phi_{jcd} \Phi_{ief} 
\psi^{ce}\psi^{df}
\nonumber\\&&
+ \half g^{jk} \Phi_{jcd} \Phi_{ike} 
\psi^{cd}t^e t_a
- g^{jk}g^{mn}\Phi_{jma}\Phi_{ikn}
\nonumber\\&&
+ \half \Phi_{icd} \Pi_{be} t_a 
                            \left(\psi^{cb}\psi^{de}
                      +\half\psi^{be} t^c t^d\right)
\nonumber\\&&
- \Phi_{icd} \Pi_{ba} t^c \left(\psi^{bd}
                            +\half t^b t^d\right)
\nonumber\\&&
+ \half \gamma_2 \left(t_a \psi^{cd}
- 2 \delta^c_a t^d\right) {\cal C}_{icd}
-\Delta^b{}_{ia}{\cal C}_b.
\end{eqnarray}
Finally, the four-index constraint, $\mathcal{C}_{ijab}$, defined
in Eq.~(\ref{e:FourIndexConstraintDef}), is given by
\begin{eqnarray}
{\cal C}_{ijab} &=& 2\tilde\nabla_{[j}\Phi_{i]ab}
+\tilde R^c{}_{aji}\psi_{cb}
+\tilde R^c{}_{bji}\psi_{ac}.
\label{e:FourIndexConstraint}
\end{eqnarray}
These expressions for the 
constraints make it possible to evaluate
them easily in terms of the first-order dynamical fields of the system
and their spatial derivatives at any instant of time.  These
expressions are analogous to those for the standard noncovariant
generalized harmonic evolution system~\cite{Lindblom2006}, but the
covariant expressions used here depend in critical ways on the
geometry of the reference metric $\tilde\psi_{ab}$ used to define the
covariant derivative $\tilde\nabla_i$ .

\section{Tensor Harmonics on $S^3$}
\label{s:AppendixB}

This appendix summarizes the basic properties of the three-sphere
scalar, vector, and tensor harmonics.  These harmonics are defined here
as eigenfunctions of the covariant Laplace operator on the
three-sphere, based on the approach developed by
Sandberg~\cite{Sandberg1978}.  The notation introduced here is
intended to be simpler and more systematic than that used by Sandberg.
Our expressions for the vector and tensor harmonics are also
covariant.  Covariance allows us to evaluate these tensors using any
convenient choice of coordinate basis on $S^3$, like the multicube
Cartesian coordinates.  The angular functions $\chi$, $\theta$, and
$\varphi$ that appear in our expressions are considered to be
functions of whatever choice of spatial coordinates is used.  Explicit
expressions for these angular functions in terms of the multicube
Cartesian coordinates are given, for example, in Appendix~A.3 of
Ref.~\cite{LindblomSzilagyi2011a}.

The scalar harmonics on the three-sphere are denoted here as $Y^{k\ell
  m}$, where $k\ge \ell\geq 0$ and $\ell\geq m \geq -\ell$ are
integers.  These harmonics are defined to be eigenfunctions of the
covariant Laplace operator for the standard round metric on $S^3$:
\begin{eqnarray}
\nabla^i\nabla_i Y^{k\ell m} = -\frac{k(k+2)}{R_3^2} Y^{k\ell m},
\end{eqnarray} 
where $\nabla_i$ is the covariant derivative, and $R_3$ is the radius
of the round-sphere metric on $S^3$.  

The vector harmonics on $S^3$ can be derived directly from the scalar
harmonics.  In particular, the three vector harmonics 
$Y^{k\ell m}_{(0\,)i}$, $Y^{k\ell m}_{(1)\,i}$, and $Y^{k\ell m}_{(2)\,i}$ are
given by
\begin{eqnarray}
Y^{k\ell m}_{(0)\,i}&=& \nabla_iY^{k\ell m},
\label{e:Yklm(0)i}\\
Y^{k\ell m}_{(1)\,i}&=& \epsilon_{i}{}^{jk}\nabla_jY^{k\ell m}\nabla_k\cos\chi,
\label{e:Yklm(1)i}\\
Y^{k\ell m}_{(2)\,i}&=& \epsilon_i{}^{jk}\nabla_jY^{k\ell m}_{(1)\,k},
\label{e:Yklm(2)i}
\end{eqnarray}
where $\epsilon_{ijk}$ is the totally antisymmetric tensor
volume element, which satisfies $\nabla_n\epsilon_{ijk}=0$.
These vector harmonics satisfy the following divergence conditions:
\begin{eqnarray}
\nabla^iY^{k\ell m}_{(0)\,i}&=& -\frac{k(k+2)}{R_3^2}Y^{k\ell m},\\
\nabla^iY^{k\ell m}_{(1)\,i}&=& 0,\\
\nabla^iY^{k\ell m}_{(2)\,i}&=& 0,
\end{eqnarray}
and the following eigenvalue equations:
\begin{eqnarray}
\nabla^j\nabla_jY^{k\ell m}_{(0)\,i}&=& \frac{2-k(k+2)}{R_3^2}Y^{k\ell m}_{(0)\,i},\\
\nabla^j\nabla_jY^{k\ell m}_{(1)\,i}&=& \frac{1-k(k+2)}{R_3^2}Y^{k\ell m}_{(1)\,i},\\
\nabla^j\nabla_jY^{k\ell m}_{(2)\,i}&=& \frac{1-k(k+2)}{R_3^2}Y^{k\ell m}_{(2)\,i}.
\end{eqnarray}

There are six (symmetric) tensor harmonics on $S^3$, 
$Y^{k\ell m}_{(0\,)ij}$, $Y^{k\ell m}_{(1)\,ij}$, $Y^{k\ell m}_{(2)\,ij}$,
$Y^{k\ell m}_{(3\,)ij}$, $Y^{k\ell m}_{(4)\,ij}$, and $Y^{k\ell m}_{(5)\,ij}$,
which can be defined in terms of the scalar and vector harmonics:
\begin{eqnarray}
&&\!\!\!\!\!\!
\!\!\!\!\!\!
Y^{k\ell m}_{(0)\,ij}= Y^{k\ell m}g_{ij},
\label{e:Yklm(0)ij}\\
&&\!\!\!\!\!\!
\!\!\!\!\!\!
Y^{k\ell m}_{(1)\,ij}= \half\left(\nabla_iY^{k\ell m}_{(1)\,j}
+\nabla_jY^{k\ell m}_{(1)\,i}\right),
\label{e:Yklm(1)ij}\\
&&\!\!\!\!\!\!
\!\!\!\!\!\!
Y^{k\ell m}_{(2)\,ij}= \half\left(\nabla_iY^{k\ell m}_{(2)\,j}
+\nabla_jY^{k\ell m}_{(2)\,i}\right),
\label{e:Yklm(2)ij}\\
&&\!\!\!\!\!\!
\!\!\!\!\!\!
Y^{k\ell m}_{(3)\,ij}= \nabla_i Y^{k\ell m}_{(0)\,j}
+\frac{k(k+2)}{3R_3^2}Y^{k\ell m}_{(0)\,ij},
\label{e:Yklm(3)ij}\\
&&
\!\!\!\!\!\!
\!\!\!\!\!\!
Y^{k\ell m}_{(4)ij}= E^{k\ell} Y^{k\ell m}_{(1)\,ij}\nonumber\\
&& - \frac{1}{4\sin^2\chi}
\left(Y^{k\ell m}_{(1)\,i}\nabla_j\cos\chi+ Y^{k\ell m}_{(1)\,j}\nabla_i\cos\chi\right)
\nonumber\\
&&\quad\times \left\{
\left[\ell(\ell+1)-2\right](E^{k\ell})^2
+6\cos\chi E^{k\ell}-4\right\},\quad
\label{e:Yklm(4)ij}\\
&&\!\!\!\!\!\!
\!\!\!\!\!\!
Y^{k\ell m}_{(5)\,ij}= \half\left(\epsilon_i{}^{sn}\nabla_sY^{k\ell m}_{(4)\,nj} 
+\epsilon_j{}^{sn}\nabla_sY^{k\ell m}_{(4)\,ni}\right) .\qquad
\label{e:Yklm(5)ij}
\end{eqnarray}
In Eq.~(\ref{e:Yklm(4)ij}), the quantity $H^{k\ell}(\chi)$
is the function that transforms $S^2$ harmonics into $S^3$
harmonics: $Y^{k\ell m}(\chi,\theta,\varphi)
=H^{k\ell}(\chi)Y^{\ell m}(\theta,\varphi)$, while $E^{k\ell}(\chi)$
is defined by
\begin{eqnarray}
\!\!\!\!\!\!
E^{k\ell}=\frac{2}{\left[2-\ell(\ell+1)\right]\sin\chi H^{k\ell}}\,
\frac{d}{d\chi}\left(\sin^2\chi H^{k\ell}\right).
\label{e:EklDef}
\end{eqnarray}

These tensor harmonics are trace free, 
$0=g^{ij}Y^{k\ell m}_{(1)\,ij}=g^{ij}Y^{k\ell m}_{(2)\,ij}=g^{ij}Y^{k\ell m}_{(3)\,ij}
=g^{ij}Y^{k\ell m}_{(4)\,ij}=g^{ij}Y^{k\ell m}_{(5)\,ij}$,
except for
$g^{ij}Y^{k\ell m}_{(0)\,ij}= 3Y^{k\ell m}$.  These tensor harmonics
satisfy the following divergence conditions:
\begin{eqnarray}
\nabla^iY^{k\ell m}_{(0)\,ij}&=& Y^{k\ell m}_{(0)\,j},\\
\nabla^iY^{k\ell m}_{(1)\,ij}&=& \frac{3-k(k+2)}{2R_3^2}Y^{k\ell m}_{(1)\,j},\\
\nabla^iY^{k\ell m}_{(2)\,ij}&=& \frac{3-k(k+2)}{2R_3^2}Y^{k\ell m}_{(2)\,j},\\
\nabla^iY^{k\ell m}_{(3)\,ij}&=& \frac{2[3-k(k+2)]}{3R_3^2}Y^{k\ell m}_{(0)\,j},\\
\nabla^iY^{k\ell m}_{(4)\,ij}&=& 0,\\
\nabla^iY^{k\ell m}_{(5)\,ij}&=& 0,
\end{eqnarray}
and the following eigenvalue equations:
\begin{eqnarray}
\nabla^n\nabla_nY^{k\ell m}_{(0)\,ij}&=& -\frac{k(k+2)}{R_3^2}Y^{k\ell m}_{(0)\,ij},\\
\nabla^n\nabla_nY^{k\ell m}_{(1)\,ij}&=& \frac{5-k(k+2)}{R_3^2}Y^{k\ell m}_{(1)\,ij},\\
\nabla^n\nabla_nY^{k\ell m}_{(2)\,ij}&=& \frac{5-k(k+2)}{R_3^2}Y^{k\ell m}_{(2)\,ij},\\
\nabla^n\nabla_nY^{k\ell m}_{(3)\,ij}&=& \frac{6-k(k+2)}{R_3^2}Y^{k\ell m}_{(3)\,ij},\\
\nabla^n\nabla_nY^{k\ell m}_{(4)\,ij}&=& \frac{2-k(k+2)}{R_3^2}Y^{k\ell m}_{(4)\,ij},\\
\nabla^n\nabla_nY^{k\ell m}_{(5)\,ij}&=& \frac{2-k(k+2)}{R_3^2}Y^{k\ell m}_{(5)\,ij}.
\end{eqnarray}
These expressions for the 
tensor harmonics are equivalent to those
given by Sandberg~\cite{Sandberg1978}.

The scalar and tensor harmonics on $S^3$ can be computed numerically
in a straightforward way.  The scalar harmonics $Y^{k\ell m}$ are
related to the standard $S^2$ harmonics $Y^{\ell m}$ by the expression
$Y^{k\ell m}=H^{k\ell}(\chi)Y^{\ell m}(\theta,\varphi)$.  The
functions $H^{k\ell}(\chi)$ can be determined numerically for $k=\ell$
and $k=\ell+1$ by the expressions
\begin{eqnarray}
H^{\ell\ell}(\chi)&=& (-1)^{\ell+1}2^{\ell}\ell\kern 0.1em!\,
\sqrt{\frac{2(\ell+1)}{\pi(2\ell+1)!}}\sin^\ell\chi,\qquad\\
H^{\ell+1\,\ell}(\chi) &=& \sqrt{2(\ell+2)}\cos\chi H^{\ell\ell}(\chi),
\end{eqnarray}
and for $k>\ell+1$ using the recursion relation
\begin{eqnarray}
&&\!\!\!\!\!\!\!\!
H^{k+2\,\ell} = 
2\cos\chi \sqrt{\frac{(k+3)(k+2)}{(k+3+\ell)(k+2-\ell)}}\,H^{k+1\,\ell} 
\qquad\nonumber\\
&&\qquad-\sqrt{\frac{(k+3)(k+2+\ell)(k+1-\ell)}
{(k+1)(k+3+\ell)(k+2-\ell)}}\,H^{k\ell}.\qquad
\end{eqnarray}
%
%
This recursion relation for $H^{k\ell}(\chi)$ is obtained from the
standard recursion relation used to determine the associated Legendre
functions~\cite{numrec_f} and the fact that $H^{k\ell}(\chi)$ is
proportional to $Q^{\ell+1/2}_{k+1/2}(\chi)/\sqrt{\sin\chi}$, where
$Q^{\ell+1/2}_{k+1/2}(\chi)$ is the associated Legendre function of
the second kind~\cite{LindblomSzilagyi2011a}.

The quantities $E^{k\ell}(\chi)$, defined in Eq.~(\ref{e:EklDef}),
can be obtained from $H^{k\ell}(\chi)$ using the standard
expressions for the derivatives of the associated Legendre
functions.  For $k=\ell$, we have
\begin{eqnarray}
\!\!\!\!\!\!
E^{\ell\ell}&=&-\frac{2\cos\chi}{\ell-1},
\end{eqnarray}
while for $k>\ell$, these are given by the recursion relation
\begin{eqnarray}
\!\!\!\!\!\!
E^{k\ell}&=&
\frac{2(k+2)\cos\chi}{2-\ell(\ell+1)} \nonumber\\
&&- \frac{2\sqrt{(k+1)(k-\ell)(k+\ell+1)}}{\left[2-\ell(\ell+1)\right]\sqrt{k}}
\frac{H^{k-1\,\ell}}{H^{k\ell}}.\qquad
\end{eqnarray}
%
\vfill\eject 

\bibstyle{prd} \bibliography{../References/References}

\end{document}